\documentclass[12pt]{iopart} 
\usepackage{graphics}
\usepackage{cite}
\usepackage{iopams}
\begin{document}
\def\ave#1{\langle#1\rangle} 
\def\vS{{\bi S}}            
\def\vT{{\bi T}}            
\def\vsigma{\boldsymbol\sigma}
\def\Stot{S_{\rm tot}}       
\def\Ms{M_{\rm s}}           
\def\cH{{\cal H}}        
\def\cHeff{{\cal H}_{\rm eff}}         
\def\Jr{J_{\rm r}}          
\def\dstyle{\displaystyle}
\def\ket#1{|#1\rangle}    
\def\uspin{\uparrow}       
\def\dspin{\downarrow}       
\jl{1}
\review
[Quantum spin nanotubes --- frustration, competing orders and 
criticalities ---]
{Quantum spin nanotubes --- frustration, competing orders and 
criticalities ---}
 \author{T\^oru Sakai$^1$, Masahiro Sato$^2$\footnote{Presend address: Department of Physics and Mathematics, Aoyama Gakuin University, 5-10-1 Fuchinobe, Chuo-ku, Sagamihara-shi, Kanagawa 251-5258, Japan}, Kiyomi Okamoto$^3$, Kouichi Okunishi$^4$ and Chigak Itoi$^5$}
 
 \address{$^1$Japan Atomic Energy Agency, 
Spring-8, 1-1-1 Kouto, Sayo, Hyogo 679-5148, Japan, 
and  Department of Material Science, University of Hyogo, Kamigori, 
Hyogo 678-1297, Japan, 
and JST TRIP, Japan}
 
 \address{$^2$Condensed Matter Theory Laboratory, RIKEN, Wako, Saitama 351-0198, Japan}
 
 \address{$^3$Department of Physics, Tokyo Institute of Technology, 2-12-1, O-okayama, Meguro-ku, Tokyo 152-8551, Japan }

 \address{$^4$Department of Physics, Niigata University, Niigata 950-2181, Japan}
 
 \address{$^5$Department of Physics, Nihon University, Kanda-Surugadai, Chiyoda-ku, Tokyo 101-8308, Japan}
\begin{abstract}
Recent developments of theoretical studies on  spin 
nanotubes are reviewed,
especially focusing on  the $S=1/2$ three-leg spin tube. 
In contrast to the three-leg spin ladder, the tube has a spin 
gap in case of the regular-triangle unit cell
when the rung interaction is sufficiently large. 
The effective theory based on the Hubbard Hamiltonian indicates a quantum phase 
transition to the gapless spin liquid due to the lattice 
distortion to an isosceles triangle. 
This is also supported by the 
numerical diagonalization and the density matrix renormalization 
group analyses. 
Furthermore, combining analytical and numerical approaches, 
we reveal several novel magnetic-field-induced phenomena: 
N\'eel, dimer, 
chiral and/or inhomogeneous orders, new mechanism for the magnetization plateau formation,
and others. 
The recently synthesized 
spin tube materials are also briefly introduced. 

\end{abstract}
\maketitle
%
%
\section{Introduction}
%
The geometrically frustrated low-dimensional quantum spin systems 
have attracted 
increasing attention in recent years. 
Among them the spin nanotube is one of the most interesting systems, because 
it is expected to be a new generation of multifunctional devices. 
Recently some candidates for the spin nanotubes have been 
synthesized; 
three-leg tubes [(CuCl$_2$tachH)$_3$Cl]Cl$_2$
\cite{seeber,nojiri} and CsCrF$_4$ \cite{manaka}, 
a nine-leg one ${\rm Na_2 V_3 O_7}$ \cite{millet}, and 
a four-leg one $\rm Cu_2Cl_4\cdot D_8C_4SO_2$ (Refs.~\cite{garlea}
and \cite{zheludev}). 
Several theoretical works indicated various exotic quantum phenomena 
of the spin tubes \cite{schulz,kawano,cabra,mila,fouet,
okunishi,okunishi2,sakai,sakai4,arikawa,sato-sakai,sato07,sato05,
sato-oshikawa,sakai2}. 
Most of them concentrated on the $S=1/2$ three-leg spin tube, because 
both of frustration and quantum fluctuation are the largest. 
In this paper, we review the recent theoretical results of the system.

In the $S=1/2$ three-leg antiferromagnetic spin tube 
the unit cell consists of three spins. 
According to the Lieb-Schultz-Mattis theorem~\cite{lieb}, 
the spin gap must be accompanied with at least doubly-degenerate 
ground states. 
In fact, previous numerical analyses~\cite{kawano,sakai,arikawa}
have confirmed such doubly-degenerate $S=0$ ground states 
due to the spontaneous breaking of the translational symmetry along the
leg direction. 
The ground states have a valence-bond type
(dimerized) order \cite{kawano,arikawa}. 
Some numerical works \cite{sakai,arikawa,sakai2} indicated that 
the spin gap is quite 
fragile against the lattice distortion changing the unit cell 
from a regular triangle to an isosceles one. 
As a result, when one of the three rung couplings is changed, 
a quantum phase transition occurs from the spin gap to gapless 
phases. 
Developing the effective theory based on the Hubbard Hamiltonian, 
using the numerical exact diagonalization and the density matrix 
renormalization group (DMRG) approach, 
the ground-state phase diagram of the quantum phase transition 
was presented \cite{sakai2}, reviewed in sections 3 and 4. 

This system was also theoretically revealed to exhibit some 
exotic phenomena in magnetic field. 
The numerical diagonalization study \cite{cabra} 
suggested that a magnetization plateau 
appears at 1/3 of the saturation for sufficiently large rung interactions. 
Introducing the above lattice distortion, 
two different plateau formation mechanisms; the up-up-down (uud) and 
the dimer-monomer types, are expected to appear,  
depending on types of isosceles triangle. 
Our recent work \cite{sakai3} 
has indicated a new plateau phase with a dimer and/or 
chiral  order accompanied with the staggered moment. 
On the other hand, the bosonization for the weak rung-coupling regime~\cite{sato07} 
has shown that a vector chiral phase is 
widely present in the field-induced Tomonaga-Luttinger liquid (TLL) phase.  
In addition, a spin-wave type approach~\cite{sato-sakai} has predicted that 
not only the chiral order but also an inhomogeneous magnetization order 
appears near the saturation. 
These field-induced phenomena will be discussed in detail in section 5. 

The valence bond solid (VBS) ground states were revealed to appear 
in some different spin tube systems. 
For example, the quantum Monte Carlo simulation study \cite{matsumoto}
suggested that several different types of the VBS ground state 
occur in some chiral spin nanotubes without frustration. 
The recent non-linear $\sigma$ model and DMRG approach \cite{charrier}
indicated some three-leg higher spin tubes exhibit several Haldane phases. 

One of the most realistic candidate of the 
$S=1/2$ three-leg spin tube is [(CuCl$_2$tachH)$_3$Cl]Cl$_2$
\cite{nojiri}. 
In this material each triangle unit cell is oriented upside-down 
to adjacent one, different from the straight spin tube (see section 6). 
The DMRG calculation of such a twisted 
spin tube \cite{fouet,okunishi,okunishi2} suggested 
that the material has no spin gap, which is consistent with 
the magnetization measurement \cite{nojiri}. 

As a future prospect, we consider the superconductivity 
in the carrier-doped spin nanotubes in section 7. 
A possible chirality induced superconductivity will be 
proposed as a new mechanism.

\section{Isosceles spin tube}

\begin{figure}[ht]
   \begin{center}
      \scalebox{0.3}{\includegraphics{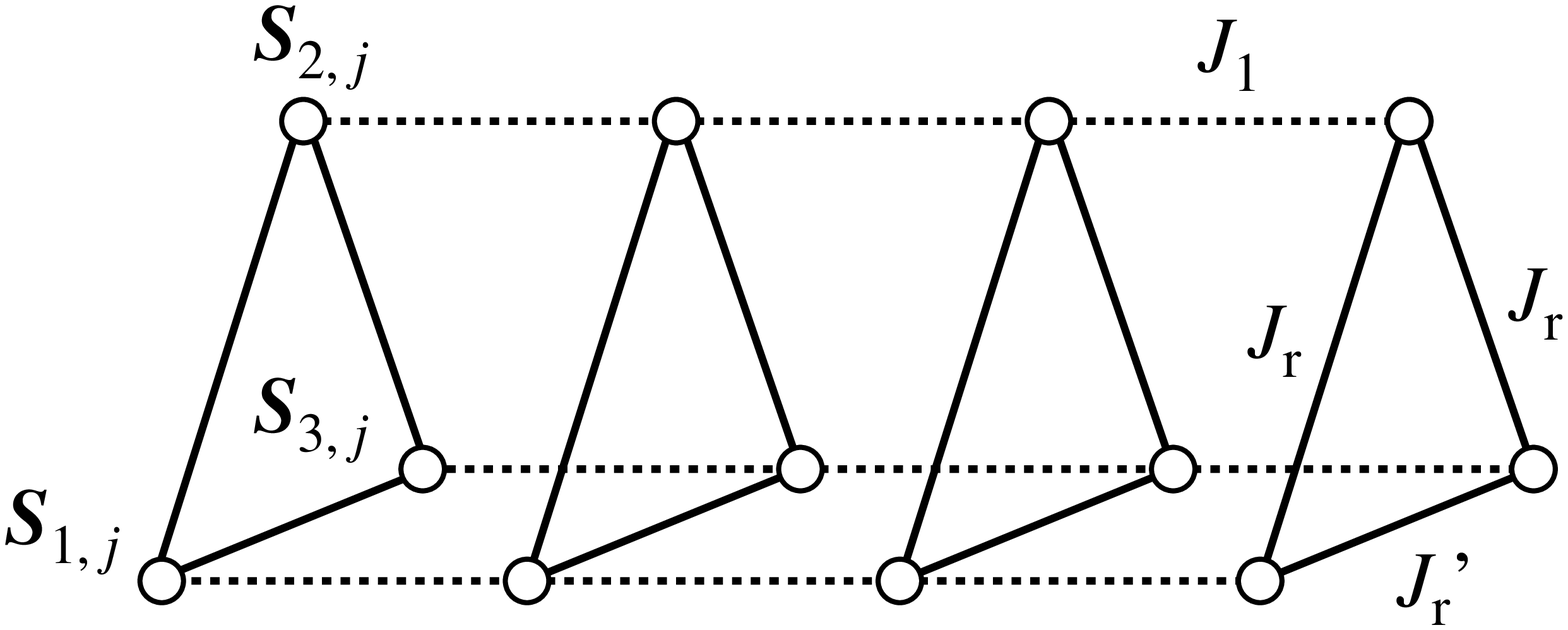}}
   \end{center}
   \caption{Sketch of the isosceles three-leg spin tube.}
   \label{fig:model}
\end{figure}

We consider the $S=1/2$ isosceles three-leg spin tube \cite{sakai2}
shown in figure ~\ref{fig:model}, 
described by the Hamiltonian 
\begin{equation}
\label{ham}
\cH= J_1 \sum _{i=1}^3 \sum_{j=1}^L \vS_{i,j}\cdot \vS_{i,j+1}
   +J_{\rm r} \sum _{i=1}^2 \sum_{j=1}^L \vS_{i,j}\cdot \vS_{i+1,j} 
   +J'_{\rm r} \sum_{j=1}^L \vS_{3,j}\cdot \vS_{1,j}
\end{equation}
where $\vS_{i,j}$ is the spin-$1/2$ operator and $L$ is the length of 
the tube along the leg direction. The exchange interaction constant $J_1$
stands for the neighboring spin pairs along the legs, while $J_{\rm r}$ and
$J'_{\rm r}$ the rung interaction constants. 
All the exchange interactions are supposed to be antiferromagnetic (namely, positive). 
The ratio $\alpha=J'_{\rm r}/J_{\rm r}$ expresses the degree of the asymmetry of
the rung interactions. 
Throughout this paper, we fix $J_{\rm r}$ to unity. 
The effect of the magnetic field $H$ along the $z$ direction is taken into account
by adding the Zeeman term $\cH_{\rm Z}=-H \sum_{i,j} S_{i,j}^z$ to (\ref{ham}). 

\section{Effective  theory}

In this section, we explain  
 low-energy effective theories for the spin tube~(\ref{ham}). 
In section~\ref{global}, we sketch a method to draw a phase diagram 
in the whole coupling-constant space $(\alpha, J_1)$ used in \cite{sakai2}.
Next, we investigate two special regimes 
$J_{\rm r} \ll J'_{\rm r}\ll J_1 $ and $J_{\rm r} \gg J_1 $, 
respectively, in sections 3.1 and 3.2.

\subsection{Global phase diagram derived 
from the Hubbard model on the tube lattice}
\label{global}
Here, we explain a systematic 
method to draw global phase diagrams of one-dimensional 
antiferromagnetic quantum spin systems.
It is well-known that $S=1/2$ Heisenberg model on an arbitrary
lattice is obtained from the corresponding half-filled Hubbard model 
in the limit of strong on-site Coulomb interactions.
Especially in one dimension, 
the spin configurations of the low-energy states
in the Heisenberg model agree with those in 
the half-filled Hubbard model even with a weak Coulomb interaction
\cite{Aff-Hal,affleck2,ledermann}. 
The phase in the weak-Coulomb regime often 
smoothly connect to those in the strong-Coulomb regime in
one-dimensional electron systems.  
By these arguments, we obtain a low-energy effective
theory for the spin tube~(\ref{ham}) from the corresponding Hubbard model. 
To discuss the wider parameter space, first
we diagonalize the kinetic parts of the Hubbard
Hamiltonian including both the leg and rung hopping
terms \cite{balents,Arrigoni,lin}. 
Then, we take account of the on-site Coulomb interaction as 
the perturbation, with help of the Non-Abelian bosonization \cite{affleck,tsvelik,gogolin}
and conformal field theory (CFT).

The Hamiltonian of the Hubbard model 
on the three-leg tube lattice
\begin{equation}
    \label{Hubbard_tube}
    \cH=\cH_{\rm hop}+\cH_{\rm int}
\end{equation}
consists of the hopping part 
\begin{eqnarray}
    \cH_{\rm hop} = \sum_{n=1} ^L \sum_{i=1}^3 \sum_{\sigma= \uparrow, \downarrow}
    (t c_{n+1,i,\sigma}^\dag c_{n,i,\sigma}+ s_{i+1,i}  
    c_{n,i+1,\sigma}^\dag c_{n,i,\sigma} \nonumber +  {\rm h.c.})
\end{eqnarray}
and the on-site interaction part
\begin{equation}
    \label{coulomb}
    \cH_{\rm int}= U \sum_{n=1} ^L \sum_{i=1}^3 n_{n,i,\uparrow} n_{n,i, \downarrow}
\end{equation}
where $n_{n,i,\sigma}=c_{n,i,\sigma}^\dag c_{n,i,\sigma}$ and $U > 0$ is
the repulsive coupling constant. The electron operators $c_{n,i,\sigma}$ 
and $c_{n,i,\sigma}^\dag$ satisfy the periodic boundary conditions for
both the leg and the rung directions,
$c_{n+L,i,\sigma}
= c_{n,i,\sigma}, \ \ \ c_{n,i+3,\sigma}=c_{n,i,\sigma},
$ 
and anticommutation relations. 
The hopping parameters are 
given by $t>0$, $s_{1,2}=s_{2,3}=s>0$ and $s_{3,1}= \beta s > 0$. 
The strong coupling expansion shows that
this model at the half-filling case is reduced to 
the Heisenberg model with $J_1=4t^2/U$, $J_{\rm r}= 4s^2/U$ and 
$\alpha = \beta^2$.

By a suitable unitary transformation, 
the hopping Hamiltonian can be written into the following diagonal form
\begin{equation}
   \cH_{\rm hop}
    = \sum_{k} \sum_{i=1}^3 \sum_{\sigma = \uparrow, \downarrow} 
      E_{i}(k)d_{k,i,\sigma}^\dag d_{k,i,\sigma}
\end{equation}
where the wave number $k$ is summed over $\frac{2 \pi}{L} \leq k \leq 2 \pi$.
With an orthogonal matrix $O_{ij}$,
the operators $d_{k,i,\sigma}$ and $d_{k,i,\sigma}^\dag$ are defined by
\begin{eqnarray}
    d_{k,i,\sigma} = 
    \frac{1}{\sqrt{L}}\sum_{n=1}^L \sum_{j=1}^3 \exp({-\rmi kn}) O_{ij} c_{n,j,\sigma},  \\
    d_{k,i,\sigma}^\dag= 
    \frac{1}{\sqrt{L}}\sum_{n=1}^L \sum_{j=1}^3 \exp({\rmi kn}) O_{ij} c_{n,j,\sigma}^\dag
\end{eqnarray}  
which satisfy the standard anticommutation relations.
The energy
eigenvalues of the one-electron states are
\begin{eqnarray}
    &&E_{1}(k)= -\beta s +2 t \cos k, \\
    &&E_{2}(k)= \frac{1}{2}(\beta s - s\sqrt{\beta^2 + 8} + 4 t \cos k), \\
    &&E_{3}(k)=  \frac{1}{2}(\beta s + s\sqrt{\beta^2 + 8} + 4 t \cos k).
\end{eqnarray}
Note that a degeneracy $E_1(k)=E_2(k)$ appears at $\beta=1$ 
due to the permutation symmetry
($\vec S_{2,j}\leftrightarrow \vec S_{i,j}$) for $i=1,2,3$.

For the half-filled case, $3L$ one-electron 
states should be occupied by electrons with up and down spins.
As a result, the ground state of the hopping Hamiltonian 
has one, two or three pairs of the Fermi points $(k_j,\bar{k}_j)$
just on the Fermi sea, depending on the parameters $s/t$ and $\beta$. 
Since the low-energy excitations are
given by the particle-hole creations around these Fermi points,
they can be represented by using the Dirac fermions, the left mover 
$\psi_{j,\sigma}(x)$ and the right one $\bar{\psi}_{j,\sigma}(x)$, 
which are defined from the electrons around the $j$-th pair of Fermi points. 
If the $j$-th band has no Fermi points in the half-filled case, 
we should neglect $\psi_{j,\sigma}(x)$ and 
$\bar{\psi}_{j,\sigma}(x)$. On this understanding, 
we approximate the original electron operators in terms of
the Dirac fermions as follows: 
\begin{equation}
   c_{n,i,\sigma} \sim \sqrt{a} \sum_{j=1}^3 O^{-1} _{ij}
   \left\{ \exp(\rmi k_j x/a) \psi_{j,\sigma}(x)
   +\exp(\rmi \bar{k}_j x/a) \bar{\psi}_{j,\sigma}(x) \right\}
\end{equation} 
where $a$ is the lattice spacing with dimension of length and
$x=a n$ is the continuous position coordinate.
We add
the on-site Coulomb interaction~(\ref{coulomb}) to
this free Dirac fermion system, and use the Non-Abelian bosonization
techniques. Following the field-theory argument in 
\cite{Aff-Hal}, we expect that
for the number of Fermi-point pairs is odd (even), 
the spin excitations are gapless (gapped) in the half-filled Hubbard
tube. In particular, in the cases of one or two Fermi-point pairs, 
we can explicitly determine whether or not a spin gap exists as follows.

First, we consider the case of one pair of Fermi points
$k_1=\frac{3\pi}{2}$ and $\bar{k}_1=\frac{\pi}{2}$. 
In this case, the interaction~(\ref{coulomb}) is approximated as the sum
of a Umklapp interaction and two marginal ones
\begin{equation}
   H_{\rm int} \sim \int dx \Big[g_1 {\cal O}_1(x)-g_2 {\cal O}_2(x)-g_3
   {\cal O}_3(x)+\cdots\Big] 
\end{equation}
where $g_{1,2,3}$ are positive coupling constants proportional to $U$.
The Umklapp term is expressed as 
\begin{equation}
    \label{Umklapp}
    {\cal O}_1(x)=\psi_{1,\uparrow}(x)^\dag
    \psi_{1,\downarrow}(x)^\dag \bar{\psi}_{1,\uparrow}(x)
    \bar{\psi}_{1,\downarrow}(x)
\end{equation}
and the marginal interaction between the U(1) charge currents 
is given by
\begin{equation}
    {\cal O}_2(x) = 
    \psi_{1}(x)^\dag  
    \psi_{1}(x) \bar{\psi}_{1}(x)^\dag \bar{\psi}_{1}(x)
\end{equation}
where $\psi_{1}={}^t(\psi_{1,\uparrow},\psi_{1,\downarrow})$. 
It is known that the bosonized form of ${\cal O}_{1,2}$ contain only 
the charge degrees of freedom and they open a charge gap 
when $g_2$ is positive. 
Then, the remaining spin degrees of freedom 
are described by the gapless level-1 SU(2) Wess-Zumino-Witten (WZW)
theory \cite{affleck,tsvelik,gogolin}. 
This phenomenon, i.e., the charge-spin separation 
is well-known in the single Hubbard chain model. 
For this WZW theory, the third interaction 
\begin{equation}
    \label{current_int}
    {\cal O}_3(x) =
    \psi_{1}(x)^\dag \frac{\bsigma}{2} 
    \psi_{1}(x) \cdot \bar{\psi}_{1}(x)^\dag 
    \frac{\bsigma}{2}\bar{\psi}_{1}(x)
\end{equation}
is known to be marginally irrelevant if $g_3>0$. 
Except  the above interactions ${\cal O}_{1,2,3}(x)$, 
there is no relevant operator with the invariance 
under the one-site translation along the leg,
\begin{equation}
    \psi_{1,\sigma}(x) \rightarrow
    \rme^{\rmi k_1} \psi_{1,\sigma}(x), \ \ \  
    \bar{\psi}_{1,\sigma}(x) \rightarrow 
    \rme^{\rmi \bar{k}_1} \bar{\psi}_{1,\sigma}(x) 
\end{equation}
as in the case of the Heisenberg chain. 
The spin excitations remain gapless. 

On the other hand, when there exist two pairs of the 
Fermi points, $(k_1,\bar{k}_1)$ and $(k_2,\bar{k}_2)$,
the spin excitations are suffered from relevant interactions. 
In this case, the spin sector in the hopping part of the Hubbard tube 
is described by a level-2 SU(2) WZW model derived from
two decoupled Dirac fermions \cite{Note1}. 
The Coulomb interaction yields several perturbations for this
theory. For example, 
the following term
\begin{equation}
    \label{relevant_level2}
    \psi_{1,\uparrow} (x)^\dag\bar{\psi}_{2,\uparrow}(x)
    \bar{\psi}_{1,\downarrow}(x)^\dag \psi_{2,\downarrow}(x)
\end{equation}
contains a relevant perturbation in the level-2 WZW model, which is
invariant under the translation
\begin{equation}
    \psi_{j ,\sigma}(x) \rightarrow
    \rme^{\rmi k_j} \psi_{j, \sigma}(x), \ \  \bar{\psi}_{j, \sigma}(x) \rightarrow
    \rme^{\rmi \bar{k}_j} \bar{\psi}_{j, \sigma}(x)
\end{equation}
Particularly for $\beta=1$, much more
relevant operators are allowed by the translational invariance
because of the coincident Fermi points $k_1=k_2$ and $\bar{k}_1=\bar{k}_2$.
Therefore, we conclude that 
any gapless spin excitation generally has no chance to survive 
except particularly rare cases [e.g., when the coupling constant of
(\ref{relevant_level2}) is zero]. 
\begin{figure}[ht]
   \begin{center}
      \scalebox{0.35}{\includegraphics{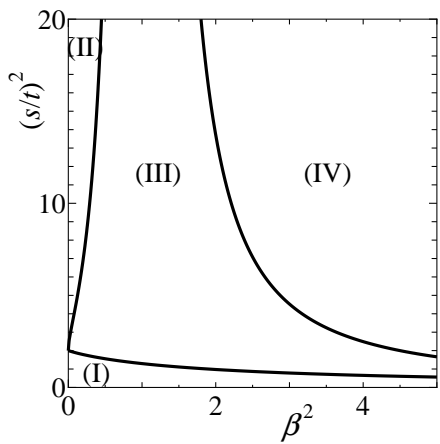}}
   \end{center}
  \caption{Phase diagram obtained from 
the effective Hubbard model~(\ref{Hubbard_tube}). 
In the strong-coupling limit ($U/t,U/s\gg 1$), the horizontal axis 
$\beta^2$ and vertical one $(s/t)^2$ can be regarded as 
$\alpha$ and $J_{\rm r}/J_1$, respectively. 
The effective theory claims that the phases (II) and (IV) are 
gapless and the phase (III) is gapful. The phase (I) is possibly gapless.}
\label{phase_eff}
\end{figure}
For the case of three Fermi-point pairs, 
the interactions among three Dirac fermions are generated from
(\ref{coulomb}),  and thus
it is difficult to analyze them and judge whether or not the spin 
excitation can survive as gapless. 
The gapless spin excitation is expected from
\cite{Aff-Hal,totsuka}. 
 
From these arguments, we can draw the ground-state phase diagram of
the half-filled Hubbard tube shown in  figure \ref{phase_eff}.
The phase (I) has three pairs of the Fermi points,
the phases (II) and (IV) have one pair, and the central phase (III) has two pairs.
Therefore, we predict that the phases (II) and (IV) have gapless spin
excitations, whereas the phase (III) possesses a spin gap. 
By $(t/s)^2=J_{\rm r}/J_1$ and $\beta^2=\alpha$ obtained from strong 
on-site interaction,
figure \ref{phase_eff} depicts the phase diagram of the $S=1/2$ three-leg spin tube~(\ref{ham}).
Since we have treated the on-site interaction
perturbatively in the effective theory, 
we should not trust the obtained 
phase boundaries as accurate ones. 
Note that the gapful phase (III) is predicted to be extended 
around the line $\beta=1$ for a finite $s/t$. In the limit 
$s/t \rightarrow\infty$, the both the left- and right-side phase
boundaries of the region (III) converge to $\beta=1$. 
This narrowing of the phase (III) is consistent with the numerical
results \cite{sakai,arikawa} 
in the strong-rung-coupling limit 
$J_{\rm r}/J_1 \rightarrow \infty$. 

Finally, we argue the universality classes 
of the phase transitions at two phase boundaries, 
(II)-(III) and (III)-(IV). 
For the level-1 SU(2) WZW model in the phases (II) and (IV), 
the most relevant perturbation is 
the marginal current-current interaction \cite{affleck,tsvelik,gogolin},
(\ref{current_int}), which is the only allowed invariant interaction.   
Since it can be marginally ``relevant'' when parameters are
finely tuned and then $g_3$ becomes negative, 
we speculate that the transition from the phases (II) or (IV) 
to (III) is caused by this marginal term. Therefore, the transitions are
expected to be in the Berezinskii-Kosterlitz-Thouless (BKT) universality
class \cite{bere,kt}.

\subsection{Gapless phase for $J_{\rm r} \ll J'_{\rm r} \ll J_1 $}
\label{gapless_phase}

In section \ref{global}, we have obtained a qualitative 
phase diagram of the spin tube~(\ref{ham}) as shown in 
figure \ref{phase_eff}. However, some subtle points still remains.
Particularly the existence of the gapless phase (I) should be discussed.
Here, we focus on the right-lower regime in figure \ref{phase_eff}
$J_{\rm r} \ll J'_{\rm r} \ll J_1, \ (\alpha =J'_r/J_{\rm r}\gg 1)$.
In this regime, we discuss the existence of the gapless phase (I)
and the phase transition between the gapless phase (I) and 
the gapful phase (III).

We introduce three level-1 SU(2) WZW models 
for the three decoupled Heisenberg chains \cite{affleck,tsvelik,gogolin}, 
and we treat their rung couplings as the perturbation 
The first and the third chains are coupled to each other 
with $J'_{\rm r}$ which is much stronger than 
two remaining couplings $J_{\rm r}$. 
It is well-known that the $J'_{\rm r}$ coupling involves a relevant 
interaction with conformal dimensions $(1/2,1/2)$ 
in the two coupled WZW models, 
it produces an energy gap \cite{shelton}. 
This relevant operator behaves like the mass operator in the 
corresponding effective field theory. 
On the other hand, like the case of one Fermi-point pair in
section \ref{global}, the WZW model for the second chain 
has the marginal irrelevant interaction
$-g_3 {\cal O}_3$
with a finite negative coupling constant $-g_3 < 0$. 
The operator ${\cal O}_3$ is equivalent to 
 (\ref{current_int}), if we use the Dirac fermions 
$(\psi_{1,\sigma},\bar\psi_{1,\sigma})$ to describe the second chain. 
The negative sign makes   $-g_3 {\cal O}_3$ irrelevant and the second
chain is gapless. A weak rung coupling $J_{\rm r}$ between this WZW model and 
the massive theory for the two coupled chains must give a correction to 
the coupling constant $-g_3$. 
If $J_{\rm r}$ is sufficiently small, the sign of 
$-g_3 < 0$ would not change and the gapless excitation is
preserved. These arguments convince us 
that the gapless phase (I) definitely exists. 
Furthermore, we expect the phase transition from the gapless phase (I)
to the gapful phase (III). Namely, if $J_{\rm r}$ exceeds a critical
value, the coupling constant $-g_3$ might change to be positive. 
In this case, the  operator $-g_3 {\cal O}_3$
becomes marginally relevant, which produces an excitation gap.
Since the operator  is marginal, 
this transition between the phases (I) and (III) 
is of the BKT
type \cite{bere,kt}, 
as in the zigzag Heisenberg chain \cite{on}.
The effective theory for
$J_{\rm r} \ll J'_{\rm r} \ll J_1, \ (\alpha =J'_r/J_{\rm r}\gg 1)$
is obtained, and the transition from the phase (I) to (III)
is discussed in \cite{sakai2}.

\subsection{Perturbation theory from the strong rung coupling limit}

In case of $\Jr \gg J_1$,
the perturbation theory from the independent triangle limit will work well.
Let us begin with the $J_1 =0$ case (i.e., three-spin problem),
which can be easily solved and there are 8 eingenstates.
Here $\psi_0^{(P)}(\Stot,\Stot^z)$ 
represents the wave function with eigenvalues $\Stot$, $\Stot^z$ and $P$,
where $P$ is the eigenvalue of the rung-parity operation $\vS_{1,j} \Leftrightarrow \vS_{3,j}$.
The state $\ket{\uspin_1 \dspin_2 \uspin_3}$
is abbreviated as $\ket{\uspin\dspin\uspin}$ for example
and $[i,j]$ denotes the singlet pair
$[i,j] = (1/\sqrt{2})(\ket{\uspin_i \dspin_j}-\ket{\dspin_i \uspin_j})$.
We show four eigenstates with $\Stot^z>0$ only,
because other four eigenstates with $\Stot^z<0$ can be easily obtained 
by interchanging $\uparrow \Leftrightarrow \downarrow$;
\begin{equation}
\fl
    \matrix{
         {\rm state}  &&{\rm energy} &S_{\rm tot}  &S^z_{\rm tot} &P\cr
     \dstyle{\psi_0^{(+)} \left({3 \over 2},+{3 \over 2}\right) = \ket{\uspin\uspin\uspin}} \hfill
            && 1/2+\alpha/4 \hfill 
            &3/2  &+3/2 &+1 \cr
     \dstyle{\psi_0^{(+)} \left({3 \over 2},+{1 \over 2}\right) = {1 \over \sqrt{3}}
                       \left( \ket{\uspin\uspin\dspin}
                              + \ket{\uspin\dspin\uspin}
                              + \ket{\dspin\uspin\uspin} \right)} \hfill 
        && 1/2+\alpha/4\hfill &3/2   &+1/2 &+1 \cr
     \dstyle{\psi_0^{(+)} \left({1 \over 2},+{1 \over 2}\right) = {1 \over \sqrt{6}}
                       \left( \ket{\uspin\uspin\dspin}
                              - 2\ket{\uspin\dspin\uspin}
                              + \ket{\dspin\uspin\uspin} \right)} \hfill
            && -1+\alpha/4 \hfill 
            &1/2  &+1/2 &+1 \cr
     \dstyle{\psi_0^{(-)} \left({1 \over 2},+{1 \over 2}\right) = {1 \over \sqrt{2}}
                       \left( \ket{\uspin\uspin\dspin}
                              - \ket{\dspin\uspin\uspin} \right)
                       = \ket\uspin_2 [1,3]       } \hfill 
        && -3\alpha/4 \hfill &1/2    &+1/2 &-1            \cr
    }~~~~
    \label{eq:psi}      
\end{equation}
These wave functions do not depend on $\alpha$
because all of them are completely classified by
the eigenvalues $\Stot$, $\Stot^z$ and $P$.
The lowest energy states are $\psi_0^{(+)}(1/2,\pm 1/2)$
for $\alpha < 1$, while they are $\psi_0^{(-)}(1/2,\pm 1/2)$
for $\alpha > 1$.

The first order perturbation theory with respect to $J_1$,
retaining four states with $\Stot=1/2$ having lower energies
and neglecting other 4 states with $\Stot=3/2$
leads to
\begin{equation}
   \cHeff
   = {J_1 \over 3}
     \sum_j \vT_j \cdot \vT_{j+1}
            \left(1 + 8 [\sigma_j^x \sigma_{j+1}^x + \sigma_j^z \sigma_{j+1}^z] \right)
     - (1-\alpha) \sum_j \sigma_j^z
     \label{eq:Heff-for-isosceles}
\end{equation}
where $\vT$ is a spin-1/2 operator acting on $\Stot^z$ and
$\sigma$ is also a spin-1/2 operator acting as
\begin{equation}
  \eqalign{
    \sigma^z \psi_0^{(\pm)} (\Stot^z ) 
    = \pm {1 \over 2} \psi_0^{(\pm)} (\Stot^z ) \cr
    \sigma^+ \psi_0^{(+)} (\Stot^z ) 
    = 0,~~~~
    \sigma^+ \psi_0^{(-)} (\Stot^z ) 
    = \psi_0^{(+)} (\Stot^z) \cr
    \sigma^- \psi_0^{(+)} (\Stot^z ) 
    = \psi_0^{(-)} (\Stot^z),~~~~
    \sigma^- \psi_0^{(-)} (\Stot^z ) 
    = 0
    }
    \label{eq:sigma}
\end{equation}
where $\Stot=1/2$ indices were omitted for simplicity.
The energy difference between $\psi_0^{(+)} (\Stot^z )$ and
$\psi_0^{(-)} (\Stot^z )$ acts as a ``magnetic field" applied on $\bsigma$ spins.
This $\cHeff$ is essentially the same as that obtained by Nishimoto and Arikawa \cite{arikawa}.

In case of sufficiently strong asymmetry $|\alpha -1| \gg 1$,
the $\bsigma$ spins are completely polarized and only $\vT$ spins survive,
where $\cHeff$ is reduced to that of the antiferromagnetic Heisenberg chain
having the gapless excitation. 
Thus the phase (III) has a finite width along the line $\Jr/J_1 = {\rm const} \gg 1$.
Next, let us focus on the region around the symmetric line $\alpha \sim 1$.
In the limit $J_1/J_{\rm r} \rightarrow 0$, 
the ground state has saturated $\bsigma$ 
spins at any asymmetric point $\alpha \neq 1$. 
For a finite $J_1/J_{\rm r}$, however, 
we expect an extended gapful phase around $\alpha=1$, if the energy gap 
exists at $\alpha=1$ due to the coupling 
between $\vT$ and $\bsigma$ spins.
The finite energy gap does not vanish by an
infinitesimal external field $\alpha-1$. 
In other words, the magnetization process 
of the $\sigma$ spins should show a zero-magnetization plateau. 
Therefore, 
an energy gap would also be present for sufficiently weak asymmetry
$|\alpha-1| \ll 1$. 

For the rest of this subsection
we suppose $\alpha=1$ (i.e. regular triangle case) 
where the translational invariance by one site along the rung direction
holds and the lowest energy states are 4-fold degenerate.
In this case another set of useful expressions for the lowest energy states is
\numparts
\begin{eqnarray}
    \ket{\uspin L}
     &=& {1 \over \sqrt{3}}
       \left( \ket{\uspin\uspin\dspin}
              + \omega \ket{\uspin\dspin\uspin}
              + \omega^{-1} \ket{\dspin\uspin\uspin}
       \right)  \\
    \ket{\uspin R}
     &=& {1 \over \sqrt{3}}
       \left( \ket{\uspin\uspin\dspin}
              + \omega^{-1} \ket{\uspin\dspin\uspin}
              + \omega \ket{\dspin\uspin\uspin}
       \right)  \\
    \ket{\dspin L}
     &=& {1 \over \sqrt{3}}
       \left( \ket{\dspin\dspin\uspin}
              + \omega \ket{\dspin\uspin\dspin}
              + \omega^{-1} \ket{\uspin\dspin\dspin}
       \right)    \\
    \ket{\dspin R}
     &=& {1 \over \sqrt{3}}
       \left( \ket{\dspin\dspin\uspin}
              + \omega^{-1} \ket{\dspin\uspin\dspin}
              + \omega \ket{\uspin\dspin\dspin}
       \right)  
\end{eqnarray}
\endnumparts
where $\omega = \exp(2\pi i/3)$ and indices $L$ and $R$
denote the wave number $k=2\pi/3$ and $k=-2\pi/3$
along the rung direction, respectively.
The first order perturbation theory with respect to $J_1$,
retaining above four states with $\Stot=1/2$ and neglecting other 4 states with $\Stot=3/2$
leads to \cite{schulz, kawano}
\begin{equation}
   \cHeff
   = {J_1 \over 3}
     \sum_j \vT_j \cdot \vT_{j+1}
            \left[1 + 4 (\tau_j^+ \tau_{j+1}^- + \tau_j^- \tau_{j+1}^+) \right]
     \label{eq:Heff-for-regular}
\end{equation}
where $\vT$ is a spin-1/2 operator acting on the first indices of the above states 
and $\tau^\pm$ are  the spin-1/2 matrices acting on the second indices as
\begin{equation}
    \matrix{
    \tau^+ \ket{L} = 0, \hfill       &\tau^-\ket{R} = \ket{L} \cr
    \tau^+ \ket{R} = \ket{L}, \hfill &\tau^-\ket{L} = 0 \hfill
    }
\end{equation}
Of course this $\cHeff$ can be obtained from (\ref{eq:Heff-for-isosceles})
through the unitary transformation and letting $\alpha=1$.

Schulz \cite{schulz} analyzed $\cHeff$ using the Jordan-Wigner transformation
and concluded that the system is gapped.
Kawano and Takahashi \cite{kawano} performed the DMRG calculation for $\cHeff$
to find that Hamiltonian $\cHeff$ has a spin gap of $0.277 J_1$.
When the next-nearest-coupling term
having a special coupling constant is added to (\ref{eq:Heff-for-regular})
they found the exact ground state closely similar to the Majumdar-Ghosh state,
which is spontaneously dimerized and breaks the translational symmetry. 
They insisted that the ground state of (\ref{eq:Heff-for-regular})
is similar to this Majumdar-Ghosh state,
and provided strong numerical evidences. 

\section{Spin gap and ground-state phase diagram --- numerical study}

In this section the numerical ground state phase diagram of the isosceles 
spin tube obtained in our previous work\cite{sakai2} is presented. 
The method to derive it is the phenomenological renormalization combined 
with the numerical diagonalization up to $L=10$ and the DMRG up to $L=128$. 

A useful order parameter to determine the phase boundaries  between the
gapless and the gapful phases
is the spin gap $\Delta$, which is the energy gap between the singlet
ground state and the triplet excited state in the finite, but large system. 
We calculate it by means of the DMRG up to $L=128$.

\begin{figure}
   \begin{center}
      \rotatebox{-90}{\scalebox{0.35}{\includegraphics{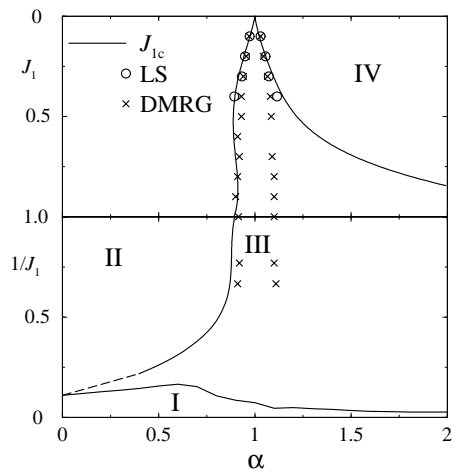}}}
   \end{center}
\caption{ 
Ground-state phase diagram of the isosceles three-leg spin
 tube (\ref{ham}), derived from the numerical analysis. 
The phases (I)-(IV) correspond to those in figure \ref{phase_eff}. 
The cross points are determined by the DMRG, 
and the solid lines by the numerical diagonalization. 
The circle points are obtained from the level-spectroscopy method 
in section 4.}
\label{phasen}
\end{figure}

In order to determine a phase boundary of the usual second order phase
transition, the phenomenological renormalization equation 
$L_1\Delta_{L_1}(\alpha_c)=L_2\Delta_{L_2}(\alpha_c)$ is often used
effectively. For the present critical point $\alpha_c$, however, 
this type of phenomenological renormalization has no clear crossing point, 
but the scaled gap $L\Delta$ increases 
with increasing $L$ in both gapless and gapful phases. 
This is because the scaled gap is an increasing function with respect to
$L$ not only in the gapful phase but also  in the gapless phase, since
the finite-size gap must have the logarithmic size correction term 
$\sim-1/\log L$. 
Here, we should recall that the logarithmic correction normally vanishes 
just at $\alpha_c$ due to the SU(2) symmetry in the $c=1$ CFT \cite{on}. 
Therefore, instead of using the crossing point of the scaled gaps, we
can estimate $\alpha_c$ as a point where the size correction is
minimized. 
Note that the minimum value $L_1\Delta_{L_1}-L_2\Delta_{L_2}$ decreases
as the size increases. This phenomenon and the assumption of the
BKT type transition suggest that this minimal value 
approaches zero as the system size increases. This is quite reasonable
if we suppose the most important finite-size correction
to the scaled gap $L \Delta$ next to $1/ \log L$ term is order of 
$1/L^2$ \cite{on,cardy2,cardy3}.
We thus determine $\alpha_c$ from the minimums for two large systems
with $L_1=96$ and $L_2=128$ for $J_1<2$. 
The estimated $\alpha_{c1}$ and $\alpha_{c2}$ are shown as crosses in
figure \ref{phasen}. 
They correspond to the phase boundary between two regions (II) and (III)
and that between (III) and (IV), respectively. 
At least these boundaries for the strong-rung-coupling regime $J_1\le 2$
are precise enough to justify that a finite gapful phase (III) exists. 
However, it is difficult to obtain $\alpha_c$ for $J_1 >2 $, because the
DMRG calculation does not well converged there. 

In order to determine the phase boundaries for the weak-rung-coupling
regime $J_1>2$, 
we use the minimum points of $L_1\Delta_{L_1}-L_2\Delta_{L_2}$
calculated by the numerical diagonalization up to $L=10$ under the
periodic boundary condition. 
Using the estimated phase boundaries for 
$(L_1,L_2)=(6,8)$ and (8,10), and 
assuming the size
correction is proportional to $1/L^2$ in both directions of $J_1$ and
$\alpha$ the phase boundaries among the phases (I), (II), (III) and (IV) 
in the thermodynamic limit were estimated. 
The phase boundaries are also shown as solid curves in figure \ref{phasen}. 
At least the phase boundaries (II)-(III) and (III)-(IV) are consistent
with the DMRG results for $J_1 \ll 1$. 
The boundary (III)-(IV) is, however, significantly deviated from the
DMRG estimation for $J_1\sim 1 $. 
This discrepancy is supposed to be due to the error of extrapolation. 
This analysis also justifies the existence of the phase (I). 
However, the error of extrapolation becomes larger as we approach the
line $1/J_1=0$ in the case of $\alpha<1$. 
Thus it is difficult to conclude that the phase (I) really exists for
$\alpha<1$ within the present numerical demonstration. 
It is also difficult to confirm the boundary (I)-(II), 
and the phase (I) might combine with the phase (II) in a certain regime
with $\alpha<1$. 

The phase boundaries between the phases (I) and (III), as well as (III) and 
(II), for smaller $J_1$ ($0<J_1<0.5$), 
were well reproduced by the level spectroscopy method \cite{on,no,LS2,LS3}, 
which is one of 
precise techniques to estimate 
the critical point for the BKT transition. 
But the level spectroscopy method does not work well for larger $J_1$,
because of too large finite-size corrections.

In our previous work \cite{sakai2}, 
applying the conformal field theory analysis \cite{cft1,cft2,cft3,GS,IK,ZS}, 
the central charge and the critical exponents of the spin correlation 
functions were estimated. 
The results suggested that all the phase boundaries in figure \ref{phasen} 
belong to the BKT universality class.

\section{Field-induced phenomena}

In this section, we consider effects of applied magnetic field in three-leg 
spin tubes. 
Particularly, we focus on field-induced phases 
(vector chiral order and magnetization plateau) and quantum phase transitions.

\subsection{Vector chiral phase in the weak rung-coupling regime}

Here, we study the weak rung-coupling regime in a magnetic 
field \cite{sato07}, namely, the Hamiltonian,
\begin{eqnarray}
    \label{eq:tube_field}
    {\cal H}_H&=&{\cal H}+{\cal H}_Z
\end{eqnarray}
with $J_1\gg |\Jr|,|\Jr'|$. 
If a sufficiently strong magnetic field is applied 
and a finite magnetization occurs, the spin-rotational SU(2) symmetry 
is reduced to the U(1) type, in which 
the Abelian bosonization \cite{gogolin,tsvelik,giamarchi}
becomes reliable at least for the weak rung-coupling regime. 
At the zero rung-coupling limit, the low-energy properties of 
$i$-th spin chain in the tube can be described by 
a free boson theory whose Hamiltonian is given by
\begin{eqnarray}
    \label{eq:TLL}
    {\cal H}_{\rm eff}^i &=& \int dx\frac{v}{2}
    \Big[K^{-1}(\partial_x\phi_i)^2+K(\partial_x\theta_i)^2\Big]
\end{eqnarray}
Here, $\phi_i(x)$ and $\theta_i(x)$ are the pair of dual scalar fields
($x=j a_0$), 
and $K$ and $v$ respectively denote the TLL parameter 
and the excitation velocity. The spin operator is also bosonized as
\begin{eqnarray}
    \label{eq:spin_boson}
    S_{i,j}^z&\approx& \frac{a_0}{\sqrt{\pi}}\partial_x\phi_i(x)
    +(-1)^j A_0 \cos(\sqrt{4\pi}\phi_i+2\pi M j)+\cdots
    \nonumber\\
    S_{i,j}^+&\approx&\exp(i\sqrt{\pi}\theta_i)
    [(-1)^j B_0+B_1\cos(\sqrt{4\pi}\phi_i+2\pi M j)+\cdots],
\end{eqnarray}
with $A_n$ and $B_n$ being nonuniversal constants. Here, 
$M=\langle S_{i,j}^z\rangle$ is the magnetization per site. 
In the present notation, $K$ runs from 
1/2 to 1 when the magnetization $M$ is increased from 0 to the 
saturated value 1/2. 
Using the formula~(\ref{eq:spin_boson}), we can obtain the bosonized 
form of the perturbative rung coupling, and the resultant effective 
Hamiltonian for the spin tube is expressed as 
\begin{eqnarray}
    \label{eq:Heff_tube}
    {\cal H}_{\rm eff} &=& \int dx\sum_{q=0}^2\frac{v}{2}
    \Big[K^{-1}(\partial_x\Phi_q)^2+K(\partial_x\Theta_q)^2\Big]
    \nonumber\\
    &&+2\sqrt{\frac{3}{\pi}}M \Jr \partial_x\Phi_0
    +\frac{M}{\sqrt{3\pi}}(\Jr-\Jr')(2\partial_x\Phi_0+\sqrt{2}\partial_x\Phi_2)
    \nonumber\\
    &&+\frac{a_0}{3\pi}(2\Jr +\Jr')(\partial_x\Phi_0)^2
    -\frac{a_0}{2\pi}\Jr'  (\partial_x\Phi_1)^2\nonumber\\
    &&-\frac{a_0}{3\pi}(2\Jr -\frac{\Jr'  }{2})(\partial_x\Phi_2)^2
    -\frac{\sqrt{2}a_0}{3\pi}(\Jr -\Jr'  )\partial_x\Phi_0\partial_x\Phi_2
    \nonumber\\
    &&+B_0^2\Jr  a_0^{-1} V(\Theta_1,\Theta_2)
    -B_0^2(\Jr -\Jr'  )a_0^{-1}\cos(\sqrt{2\pi}\Theta_1)
    \nonumber\\ 
    &&+\frac{A_0^2}{2}\Jr  a_0^{-1} V(2\Phi_1,2\Phi_2)
    -\frac{A_0^2}{2}(\Jr -\Jr'  )a_0^{-1}\cos(2\sqrt{2\pi}\Phi_1)+\cdots.
\end{eqnarray}
Here we have neglected all the terms with oscillating factors 
$\exp(\rmi\gamma M\pi j)$ or $(-1)^j$, and have introduced new pairs of 
boson fields; 
$(\Phi_0,\Theta_0)=(\sum_{i=1}^3\phi_i,\sum_{i=1}^3\theta_i)/\sqrt{3}$, 
$(\Phi_1,\Theta_1)=(\phi_1-\phi_3,\theta_1-\theta_3)/\sqrt{2}$, and 
$(\Phi_2,\Theta_2)=(\phi_1+\phi_3-2\phi_2,\theta_1+\theta_3-2\theta_2)
/\sqrt{6}$. The first line in (\ref{eq:Heff_tube}) is equivalent to 
three copies of TLLs in decoupled chains. The second line 
is linear terms which induce just a small correction to the magnetization 
$M$. 
In the isosceles case $\Jr \neq \Jr'  $, 
$\langle S_{1,j}^z\rangle=\langle S_{3,j}^z\rangle\neq 
\langle S_{2,j}^z\rangle$ would be realized. The third and fourth lines 
are quadratic terms changing the values of $K$ and $v$. In the 
regular-triangle case, the quadratic part is diagonalized in the basis
of $(\Phi_{0,1,2},\Theta_{0,1,2})$, and resultant TLL parameters and 
velocities for $(\Phi_{0,1,2},\Theta_{0,1,2})$ sectors, $K_{0,1,2}$ and 
$v_{0,1,2}$, are calculated as $K_0=K(1+\frac{2K\Jr a_0}{\pi v})^{-1/2}$, 
$K_1=K_2=K(1-\frac{K\Jr a_0}{\pi v})^{-1/2}=K_g$, 
$v_0=v(1+\frac{2K\Jr a_0}{\pi v})^{1/2}$ and 
$v_1=v_2=v(1-\frac{K\Jr a_0}{\pi v})^{1/2}=v_g$. 
The fifth and sixth lines correspond to the vertex-type perturbations,
and the function $V(\epsilon_1,\epsilon_2)$ is defined by  
\begin{eqnarray}
    \label{eq:potential}
    V(\epsilon_1,\epsilon_2)&=&2\cos\left(\sqrt{\pi/2}\epsilon_1\right)
    \cos\left(\sqrt{3\pi/2}\epsilon_2\right)
    +\cos\left(\sqrt{2\pi}\epsilon_1\right)
\end{eqnarray}
In our notation, the scaling dimension of vertex operators are given 
as $[\exp(\rmi n\sqrt{4\pi}\phi_i)]=n^2K$ and 
$[\exp(\rmi n\sqrt{\pi}\theta_i)]=n^2/(4K)$ at the decoupled case. 
We should emphasize that (\ref{eq:Heff_tube}) does not include 
any relevant terms with $\Phi_0$ and $\Theta_0$ except for commensurate 
cases with $M=q/p$ ($q$ and $p$: integer). This property is protected 
by $U(1)$ spin-rotational and translational symmetries of the spin tube. 
Therefore, the $(\Phi_0,\Theta_0)$ sector is described by a TLL. 
On the other hand, the remaining sectors are subject to the 
vertex terms.

\begin{figure}[ht]
   \begin{center}
      \scalebox{0.7}{\includegraphics{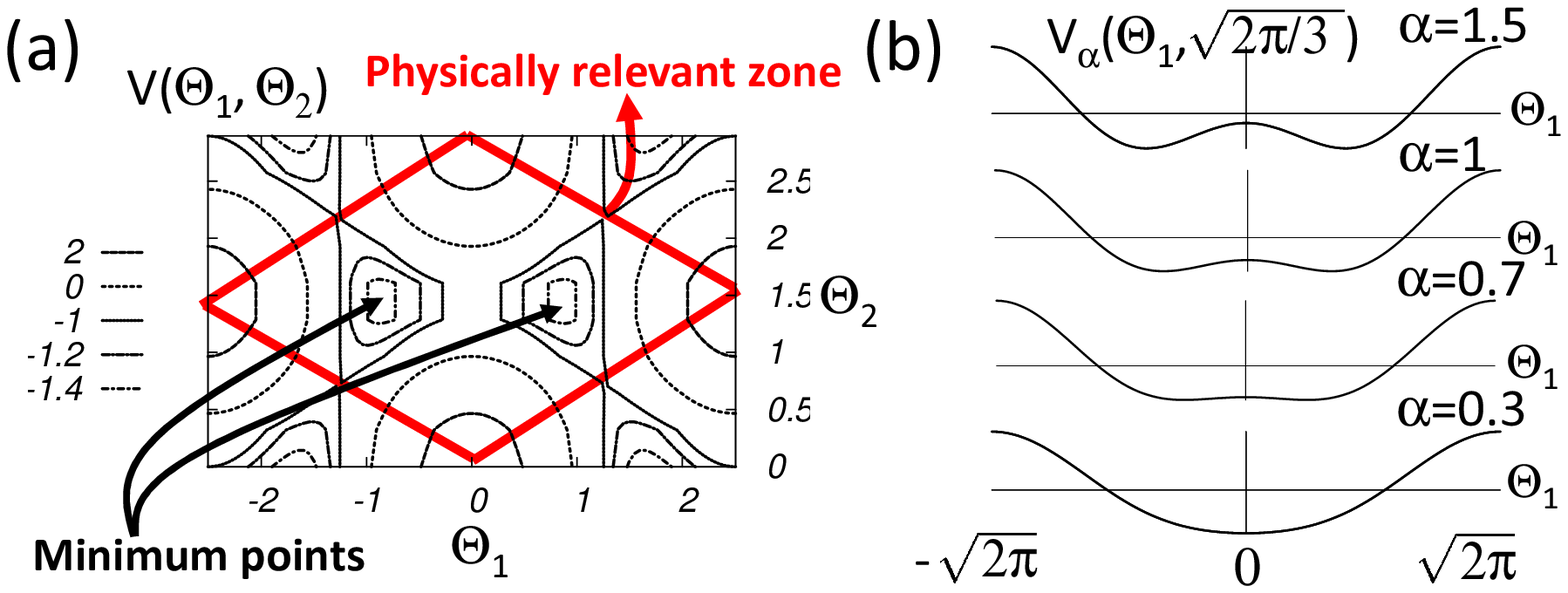}}
   \end{center}
   \caption{(a) (color online) Contour lines of potential $V(\Theta_1,\Theta_2)$ 
           at the regular-triangle case, 
           and (b) potential $V_\alpha(\Theta_1,\sqrt{2\pi/3})$ in the isosceles case 
           with several values of $\alpha$.} 
   \label{fig:potential}
\end{figure}

We first focus on the regular-triangle case $\Jr =\Jr'  $ analyzing the 
above effective Hamiltonian~(\ref{eq:Heff_tube}). 
In this case, two vertex perturbations $\cos(\sqrt{2\pi}\Theta_1)$ and
$\cos(2\sqrt{2\pi}\Phi_1)$ in (\ref{eq:Heff_tube}) vanish, and 
the scaling dimensions of $V(\Theta_1,\Theta_2)$ 
and $V(2\Phi_1,2\Phi_2)$ are readily evaluated: 
$[V(\Theta_1,\Theta_2)]=1/(2K_g)$ and $[V(2\Phi_1,2\Phi_2)]=2K_g$.
In the weak rung-coupling regime, $K_g$ is very close to $K$ 
and $K>1/2$ holds in the magnetization process. Therefore, 
$V(\Theta_1,\Theta_2)$ is more relevant and 
$(\Phi_{1,2},\Theta_{1,2})$ sectors have gapped spectra. 
The form of the potential $V(\Theta_1,\Theta_2)$ is shown in 
figure \ref{fig:potential}(a), in which the diamond region is meaningful 
in the full $\Theta_1$-$\Theta_2$ plane and it is called here 
physically relevant zone. Remarkably there are two minimum points 
in the diamond zone: 
$(\Theta_1,\Theta_2)=(\pm\sqrt{2\pi}/3,\sqrt{2\pi/3})\equiv X_\pm$. 
A minimum is mapped to the other via sign change $\Theta_1\to-\Theta_1$ 
that can be realized by the exchange of two chains 
$\vS_{1,j}\leftrightarrow \vS_{3,j}$ and 
$\theta_1\leftrightarrow \theta_3$. This suggests a spontaneous breakdown 
of rung-parity symmetry. Vector chiralities in rung bond 
$\kappa^\alpha_{i,j}=(\vS_{i,j}\times\vS_{i+1,j})^\alpha$ is a natural 
candidate for the order parameter, and the bosonized forms of 
their $z$ components are given as 
\begin{eqnarray}
    \label{eq:chirality}
    \kappa^z_{1,j}&\approx&-B_0^2 
    \sin\left(\sqrt{\pi/2}\Theta_1+\sqrt{3\pi/2}\Theta_2\right)+\cdots
    \nonumber\\
    \kappa^z_{2,j}&\approx&-B_0^2 
    \sin\left(\sqrt{\pi/2}\Theta_1-\sqrt{3\pi/2}\Theta_2\right)+\cdots
    \nonumber\\
    \kappa^z_{2,j}&\approx&B_0^2 
    \sin\left(\sqrt{2\pi}\Theta_1\right)+\cdots.
\end{eqnarray}
At the point $X_+$ ($X_-$), $\kappa_{i,j}^z$ becomes positive (negative). 
From the translational symmetry along the rung, 
we can predict $\langle\kappa^z_{1,j}\rangle
=\langle\kappa^z_{2,j}\rangle=\langle\kappa^z_{3,j}\rangle$. 
We thus conclude that in the weak rung-coupling regime with 
$\Jr =\Jr'  $ and a finite $M$, the low-energy physics is governed by 
the TLL in the $(\Phi_0,\Theta_0)$ sector and the 
remaining gapped sectors generate a vector chiral long-range order 
with spontaneously breaking the rung-parity symmetry. 
In the commensurate case of $M=1/3$, 
there appears the magnetization plateau at least in a 
strong rung-coupling regime (see the next subsection). 
It has been predicted in \cite{cabra} that 
the plateau survives up to a fairly weak rung-coupling regime 
($\Jr \sim 0.1 J_1$) at the regular-triangle case $\Jr =\Jr'  $. 
From the bosonization viewpoint, the plateau is attributed to 
the emergence of a spin gap of $(\Phi_0,\Theta_0)$ sector 
induced by a higher-order perturbation 
$\cos(\sqrt{12\pi}\Phi_0(x)+6\pi Mj)$. In fact, 
this vertex term may be relevant at $M=1/3$ in which 
the factor $\exp(\pm \rmi 6\pi Mj)$ is equal to unity. 
This corresponds to so-called gapped chiral phase; 
the system has a gapped spectrum and a vector chirality is 
long-range ordered. In other words, the spin tube offers a unique 
chance to observe both gapless and gapped vector-chiral phases 
in the magnetization process.

\begin{figure}[ht]
   \begin{center}
      \scalebox{0.7}{\includegraphics{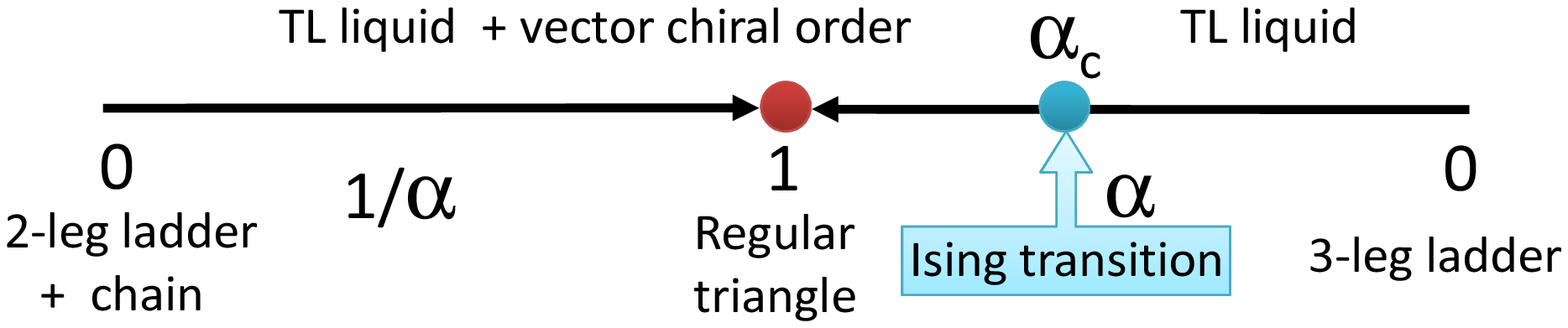}}
   \end{center}
\caption{(color online) Ground-state phase diagram of the isosceles spin tube 
in a magnetic field in weak rung-coupling regime $J_1\gg |\Jr |,|\Jr'|$.} 
\label{fig:diagram_weakrung}
\end{figure}

Next we consider effects of the rung deformation, $\Jr \neq \Jr'  $.  
Even in this isosceles case, the chirality between the first and third
chains $\langle\kappa_{3,j}\rangle$ is still useful as a order parameter 
because the tube is invariant under rung parity operation between 
these two chains. Due to the rung deformation, all kinds of the 
rung coupling terms in (\ref{eq:Heff_tube}) are modified. 
As already mentioned, the boson linear terms just slightly change
the uniform magnetization $M$. The quadratic term 
$\partial_x\Phi_0\partial_x\Phi_2$ cannot be diagonalized in the present 
basis, but its effects would be quite small and 
the qualitative nature of the vector chiral phase is expected not to 
be changed. The most important point of the rung deformation 
is the change of the vertex potential due to an additional term 
$\cos(\sqrt{2\pi}\Theta_1)$ in the fifth line of (\ref{eq:Heff_tube}).
The modified potential is expressed as follows:
\begin{eqnarray}
    \label{eq:potential_modify}
    V_\alpha(\Theta_1,\Theta_2)&=&V(\Theta_1,\Theta_2)
    -\frac{\Jr -\Jr'  }{\Jr }\cos\left(\sqrt{2\pi}\Theta_1\right)
    \nonumber\\
    &=& 2\cos\left(\sqrt{\pi/2}\Theta_1\right)
    \cos\left(\sqrt{3\pi/2}\Theta_2\right)
    +\alpha\cos\left(\sqrt{2\pi}\Theta_1\right).
\end{eqnarray}
Figure \ref{fig:potential}(b) presents this potential as a function of 
$\Theta_1$ with fixing $\Theta_2=\sqrt{2\pi/3}$. 
It clearly shows that the potential form is changed from a double-well type 
to a single-well type for $\alpha<1$, while the form is always 
double-well type for $\alpha>1$. This classical argument predicts 
the critical point $\alpha_c=0.5$, and the chiral order is expected 
to disappear in $\alpha<\alpha_c$. To gain deeper understanding 
of this expectation, we consider the effective theory for 
$(\Phi_1,\Theta_1)$ sector. If we naively replace 
$\cos(\sqrt{3\pi/2}\Theta_2)$ with its mean value 
$C_2=\langle\cos(\sqrt{3\pi/2}\Theta_2)\rangle$ in the potential 
$V_\alpha(\Theta_1,\Theta_2)$, 
the effective Hamiltonian is obtained as 
\begin{eqnarray}
    \label{eq:Heff_Phi1Theta1}
    {\cal H}_{\rm eff}^{[\Phi_1,\Theta_1]} &=& \int dx\frac{v_1}{2}
    \Big[K_1^{-1}(\partial_x\Phi_1)^2+K_1(\partial_x\Theta_1)^2\Big]
    \nonumber\\
    &&+B_0^2\Jr  a_0^{-1} \left[2C_2\cos(\sqrt{\pi/2}\Theta_1)
    +\alpha \cos(\sqrt{2\pi}\Theta_1)\right] +\cdots.
\end{eqnarray}
This is nothing but a double-frequency sine-Gordon model. 
Under the condition $K_1>1/4$, this model is believed to 
exhibit an Ising-type quantum phase transition 
by tuning the ratio of two coupling constants $\alpha/(2C_2)$ 
\cite{fabrizio,bajnok}. 
We can hence predict the phase diagram of the isosceles spin tube 
with a weak rung coupling and a finite magnetization $M\neq 1/3$
as shown in  figure \ref{fig:diagram_weakrung}. 
The chiral ordering occurs in the region $0<\alpha<1$. 
The true critical value $\alpha_c$ must deviate from its classical one
$0.5$. The order parameter $\langle\kappa_{3,j}\rangle$ behaves as 
$\sim(\alpha-\alpha_c)^{1/8}$ near the transition point. 
From this figure, one also finds that once a coupling between a 
two-leg ladder and a single chain is introduced, 
a vector chiral order immediately emerges.

Similar scenarios of a vector chiral order and an Ising transition are 
also expected in other tubes consisting of spin chains with a 
TLL parameter $K>1/2$. 
For instance, even in the zero-field case, 
spin tubes with easy-plane XXZ anisotropy must 
yield a chiral order in the weak rung-coupling regime. It is known that 
the field-induced TLL phase satisfies $K>1$ in a spin-1 
antiferromagnetic Heisenberg chain \cite{fath} 
and the leading term of bosonized spin operator $S_j^+$ has the 
same form as that of (\ref{eq:spin_boson}) \cite{essler04,sato06}. 
Therefore, three-leg spin-1 tubes have a vector chiral order 
at least in high-field and weak rung-coupling regime. 
In the vicinity of the saturation field, three-leg spin-$S$ tubes can 
be analyzed by spin-wave type approach \cite{sato-sakai}. 
It also predicts the emergence of the vector chiral order in a 
certain weak rung-coupling regime. These results obviously suggest 
that a vector chiral order is generally induced by applied magnetic field 
in three-leg spin-$S$ antiferromagnetic tubes with a weak rung coupling. 
The spin-wave approach also points out the possibility of 
an inhomogeneous magnetization phase with 
$\langle S^z_{1,j}\rangle=\langle S^z_{3,j}\rangle
\neq \langle S^z_{2,j}\rangle$ \cite{sato-sakai} in the regular-triangle case.

\subsection{1/3 plateau}

The magnetization plateau at $M = M_{\rm s}/3$ surely exists for $J_1/ \Jr \to 0$
since the triangle unit cell is composed of three $S=1/2$ spins,
while it does not exist for the $J_1 / \Jr \to \infty$ 
because the system is reduced to three independent $S=1/2$ spin chains in this limit.
Thus the $M_{\rm s}/3$ plateau phase diagram on the 
$J_1 - \alpha$ plane is of immense interest.

\begin{figure}[ht]
   \begin{center}
      \scalebox{0.3}{\includegraphics{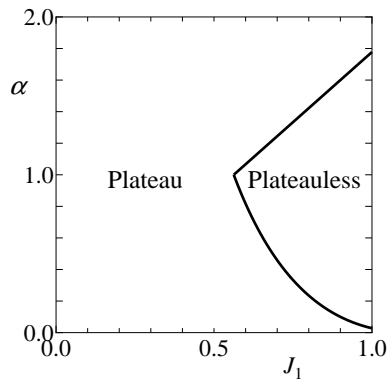}}
   \end{center}
  \caption{
The phase diagram of the 1/3-magnetization plateau predicted
by the effective Hubbard model
in the strong-coupling limit $U/t,U/s\gg 1$,
where $J_1 = (t/s)^2$ and
$\alpha = \beta^2$, respectively.}
\label{phase_eff_plateau}
\end{figure}

As discussed in section \ref{global}, we study
magnetization plateaux in
the effective Hubbard model under external field. 
The Zeeman term for electrons is given by
\begin{eqnarray}
    \cH_{\rm Z}=- H/2 \sum_{n=1}^L \sum_{i=1}^3(
    c_{n,i,\uparrow}^\dag c_{n,i,\uparrow} - 
    c_{n,i,\downarrow}^\dag c_{n,i,\downarrow} ),
\end{eqnarray}
which should be added to the Hubbard Hamiltonian.
Elaborated analysis of the effective theory for this Hubbard model 
gives the conditions for the existence of the spin excitation gap.
Then,  we
show  the existence of the 1/3 magnetization plateau and 
nonexistence of any other nonzero magnetization plateaux. Details will
published elsewhere \cite{okamotoetal}.
The phase diagram of the 1/3 magnetization plateau is depicted in
figure \ref{phase_eff_plateau}.
This agrees with the numerical calculation qualitatively,
as shown later. 

Let us discuss the 1/3 plateau problem by use of the effective Hamiltonian
(\ref{eq:Heff-for-isosceles}) supposing $\Jr \gg J_1$. 
In the 1/3 plateau case we can fix $S^z = +1/2$ and retain
the degree of freedom with respect to $\bsigma$,
resulting in
\begin{equation}
   \cHeff^{(1/3)}
   = {2J_1 \over 3}
     \sum_j \left( \sigma_j^x \sigma_{j+1}^x + \sigma_j^z \sigma_{j+1}^z \right)
     - (1-\alpha) \sum_j \sigma_j^z
     \label{eq:Heff-for-isosceles-1/3}
\end{equation}
In other words, we take only two states $\psi^{\pm}(1/2,+1/2)$ in (\ref{eq:psi}) into consideration.
We note that $\cHeff^{(1/3)}$ describes $M=\Ms/3$ states only,
because both of $\psi^{(\pm)}(1/2,+1/2)$ have $\Stot^z=+1/2$ 
corresponding to $M=\Ms/3$.

$\cHeff^{(1/3)}$ is nothing but the $S=1/2$ $XY$ chain in transverse magnetic field,
which is a special case of the $S=1/2$ $XXZ$ chain in transverse magnetic field
of which ground state was investigated by Dmitriev et al \cite{dmitriev1,dmitriev2,dmitriev3},
and Capraro and Gros \cite{capraro}.
Their results can read as follows in our cases.
Both of $\ave{\sigma_i^z \sigma_j^z}$ and $(-1)^{i-j} \ave{\sigma_i^x \sigma_j^x}$ have the long-range order
for the weak ``magnetic field" case (i. e. $0< |1-\alpha| \ll J_1$),
while only $\ave{\sigma_i^z \sigma_j^z}$ has the long-range order for
the strong ``magnetic field" case (i. e. $ |1-\alpha| \gg J_1$).
The boundary between above two cases will be approximately given by $|1-\alpha| \simeq J_1$.
In the latter case
the state of the unit triangle is essentially either $\psi_0^{(+)}(1/2,+1/2)$ or $\psi_0^{(-)}(1/2,+1/2)$
according as $\alpha <1$ or $\alpha > 1$.
In the former case, on the other hand, 
there exists the long-range N\'eel order of $\sigma$ spins along the $x$ direction
associated with the spontaneous breaking of the translational symmetry (SBTS)
and the two-fold degeneracy of the ground state. 
In this case the ground state energy
(the energy of the $\Ms/3$ state in the original spin language)
will be considerably lowered by the interaction effects,
which leads to the remarkable increase of the width of $\Ms/3$ plateau.
Further the SBTS of the $\sigma$ system
results in the SBTS and the spontaneous breaking of the inversion symmetry of
$\ave{S_j^z}$ in the original spin representation.
If we suppose the complete N\'eel order of $\sigma$ along the $x$ direction 
$\ave{\sigma_{2j}^x} = +1/2$, $\ave{\sigma_{2j+1}^x} =-1/2$ for simplicity
(this situation will be a good approximation of the ground state
in case of extremely weak ``magnetic field" $0 < |1-\alpha| \ll J_1$),
the expectation values of $S^z$ are
\begin{equation}
  \eqalign{
     &\ave{S_{1,2j}^z} =\ave{S_{3,2j+1}^z} = 0.455 \cr
     &\ave{S_{3,2j}^z} =\ave{S_{1,2j+1}^z} = 0.167 \cr
     &\ave{S_{2,2j}^z} =\ave{S_{2,2j+1}^z} = -0.122
  }
  \label{eq:Sz-analytic}
\end{equation}
which shows the spontaneous breaking of the translational symmetry ($j \Rightarrow j+1$)
and inversion symmetry ($S_{1,j} \Leftrightarrow S_{3,j}$).

\begin{figure}[ht]
   \begin{center}
      \scalebox{0.3}{\includegraphics{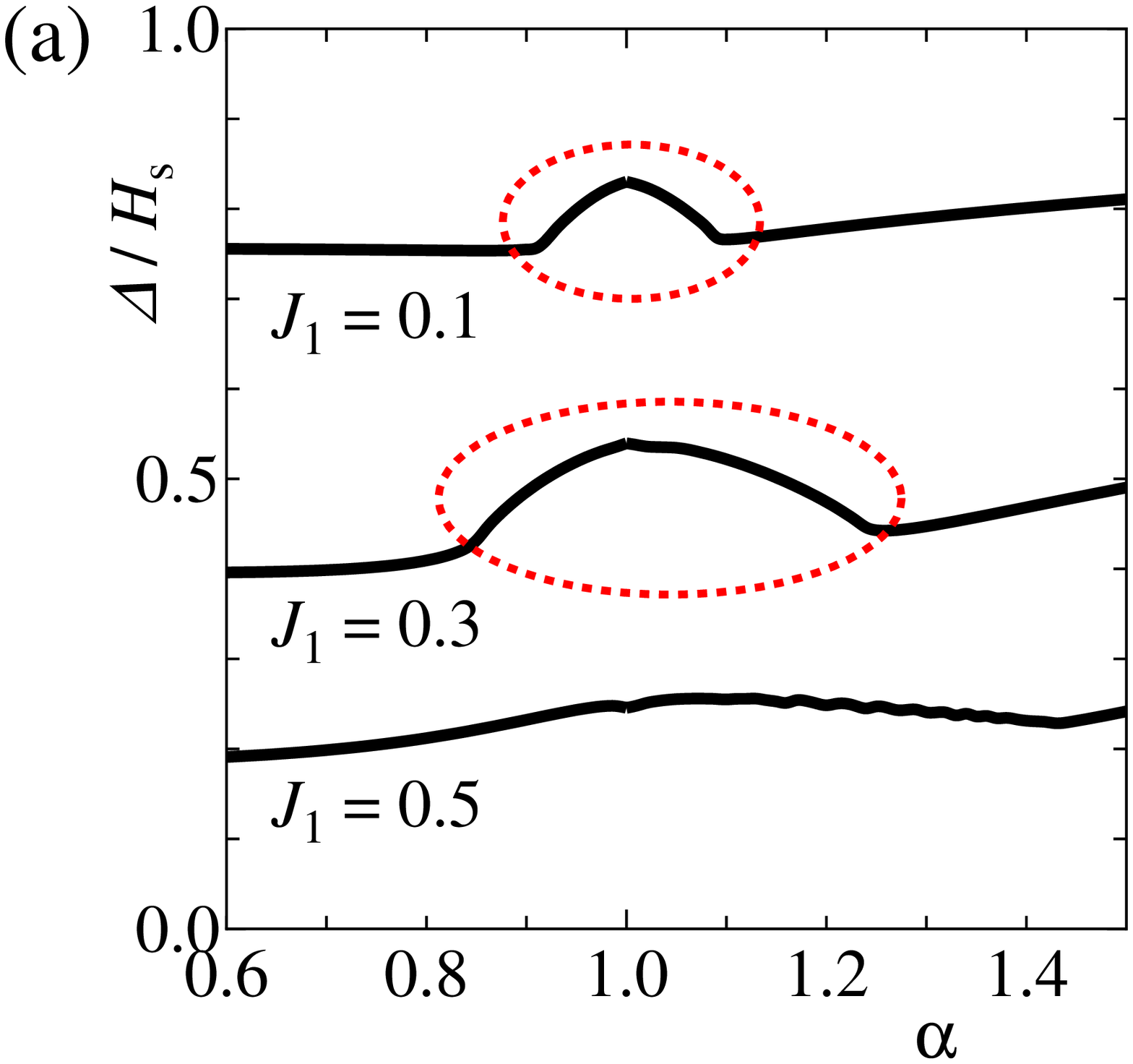}}~~~~~
      \scalebox{0.3}{\includegraphics{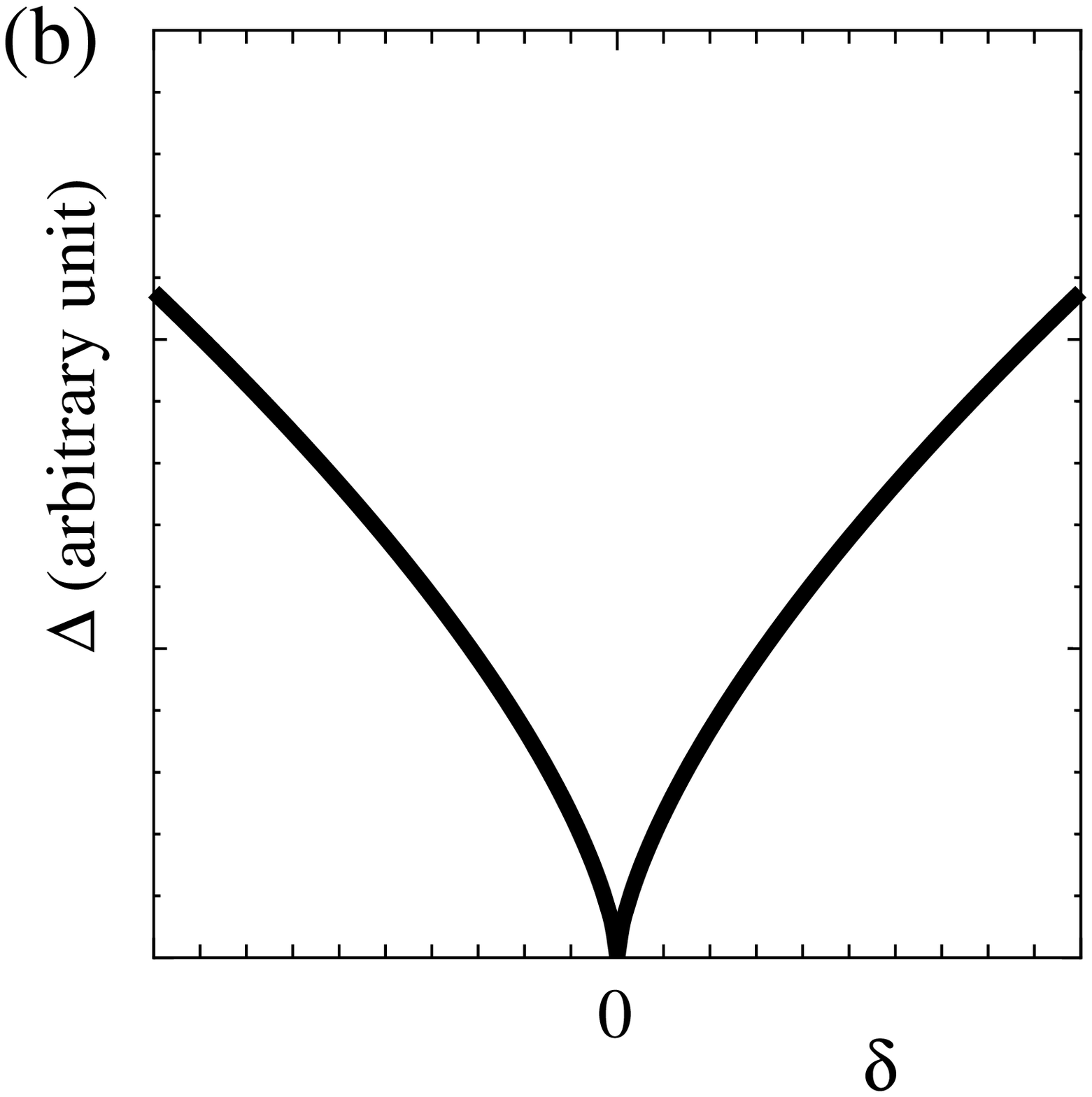}}
   \end{center}
   \caption{ (a) (color online) $\alpha$ dependence of the $\Ms/3$ magnetization plateau width
            normalized by the saturation field $H_{\rm s}$ for $J_1 = 0.1,\,0.3,\,0.5$ cases
            obtained by DMRG with $L=128$.
            The anomalous increase regions of $\Delta$ are shown by ellipses with red broken lines.
             (b) Behavior of the spin gap of the bond-alternating $S=1/2$ 
            Heisenberg chain described by the Hamiltonian (\ref{eq:bond-alt}).
            }
   \label{fig:width}
\end{figure}

Figure \ref{fig:width}(a) shows the $\alpha$ dependence
of the $\Ms/3$ magnetization plateau width $\Delta$,
from which we can see the anomalous increase of the $\Delta$ near $\alpha =1$ at least for $J_1=0.1$ and 0.3
cases as predicted by the above theoretical consideration.
In usual cases the plateau width (or the spin gap) is remarkably decreases
near the mechanism-changing point and often  becomes zero at that point,
as shown in figure \ref{fig:width}(b).
Thus, we find a new and exotic behavior of the plateau width,
which is completely opposite to that of usual cases \cite{tit-gcoe}. 
In the $J_1=0.1$ case, for instance,
the plateau is realized by the $\psi_0^{(+)}(1/2,+1/2)$ and $\psi_0^{(-)}(1/2,+1/2)$ mechanisms
for $\alpha < 0.9$ and  $1.1< \alpha$, respectively,
while by the cooperative  effects of both mechanisms for $0.9 < \alpha < 1.1$.
We note that the excitation gap (i. e., magnetization plateau width) $\Delta$
from the $M = \Ms/3$ state to $M = \Ms/3 \pm 1$ states is always finite when $\alpha$ is varied.
When $J_1 = 0.1$ and $\alpha = 0.95$, the expectation values $\ave{S^z}$ obtained by DMRG are
\begin{equation}
  \eqalign{
     &\ave{S_{1,2j}^z} =\ave{S_{3,2j+1}^z} = 0.46 \cr
     &\ave{S_{3,2j}^z} =\ave{S_{1,2j+1}^z} = 0.08 \cr
     &\ave{S_{2,2j}^z} =\ave{S_{2,2j+1}^z} = -0.04
  }
  \label{eq:Sz-DMRG}
\end{equation}
which well agree with the theoretically predicted values in (\ref{eq:Sz-analytic}).
Details will be published elsewhere \cite{okamotoetal}.

Let us explain why the new and exotic behavior of the plateau width is realized in our model,
by comparing our model with the $S=1/2$ bond-alternating Heisenberg chain described by
\begin{equation}
    \cH_{\rm b-a}
    = J \sum_j \{ 1 + (-1)^j \delta \} \vS_j \cdot \vS_{j+1}
    \label{eq:bond-alt}
\end{equation}
where $\delta$ is the bond alternation parameter.
Any small amount of $\delta$ produces the dimer spin gap as
$\Delta \propto |\delta|^{2/3}$ \cite{cross-fisher, nakano-fukuyama, ont}
(with leading logarithmic correction \cite{black-emery}).
Thus the spin gap $\Delta$ of (\ref{eq:bond-alt})
behaves as the sketch in figure \ref{fig:width}(b) near $\delta=0$.
When $\delta>0$ ($\delta < 0$), the spins $\vS_{2n}$ and $\vS_{2n+1}$
($\vS_{2n-1}$ and $\vS_{2n}$) effectively form a singlet dimer pair,
which brings about the dimer spin gap.
These two mechanism, $\delta>0$ and $\delta<0$ mechanisms,
are completely competing
and the reconstruction of the unit cell occurs at the mechanism changing point $\delta=0$.
At $\delta =0$ the system is of the Tomonaga-Luttinger state
which is characterized by the gapless excitation.
In our case, on the other hand, the unit cell is always the isosceles triangle
for both of $\alpha<1$ and $\alpha>1$ cases.
Then the reconstruction of the unit cell never occur in our case,
which enables the cooperation of two plateau-formation mechanisms,
$\psi^{(-)}(1/2,+1/2)$ mechanism and $\psi^{(+)}(1/2,+1/2)$ mechanism.

\begin{figure}[ht]
   \begin{center}
      \rotatebox{-90}{\scalebox{0.3}{\includegraphics{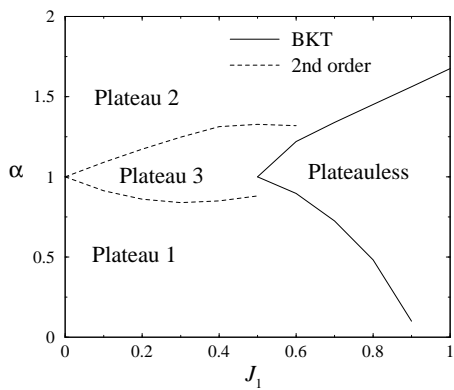}}}
   \end{center}
   \caption{Phase diagram of the $\Ms/3$ magnetization plateau.
            The plateau 1, 2 and 3 correspond to the $\psi^{(+)}(1/2,+1/2)$,
            the $\psi^{(-)}(1/2,+1/2)$, and the new plateau phases, respectively.
            Dashed lines are the second-order boundaries, 
            while solid curves are expected to be of
            the Berezinskii-Kosterlitz-Thouless (BKT) transition.}
   \label{fig:plateau-phase}
\end{figure}

The phase boundary between the traditional plateau phases and the new plateau phase
where the plateau width remarkably increases can be determined by the phenomenological renormalization
equation for the scaled gap $L\Delta_0$ obtained by the numerically exact diagonalization method
\begin{equation}
    (L + 2)\Delta_{0,L+2}(\alpha_{{\rm c},L})
     = L \Delta_{0,L}(\alpha_{{\rm c},L})
\end{equation}
where $\Delta_0$ is the excitation gap within the $\Ms/3$ space
and $\alpha_{{\rm c},L}$ is the size-dependent fixed point of $\alpha$.
The plateau-plateauless phase boundary can be estimated also by the
phenomenological renormalization equation where $\Delta_0$ is replaced by
$\Delta$ (plateau width) as in case of the spin gap at $M=0$.
Figure \ref{fig:plateau-phase} shows our preliminary results on the phase diagram at $\Ms/3$.
This phase diagram qualitatively agrees with figure \ref{phase_eff_plateau}.
However more detailed analyses will be necessary to draw a precise phase diagram.

\section{Twisted tube}

\begin{figure}[h]
\begin{center}
\scalebox{1.0}{\includegraphics{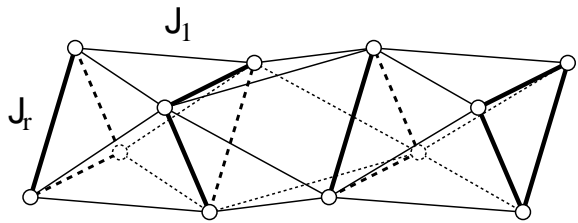}}
\end{center}
\caption{Twisted quantum spin tube. $J_{\rm r}$ indicates the exchange coupling  in the unit triangle and  $J_1$ does inter triangle coupling. 
The expansion of the twisted spin tube has the triangular lattice structure.}
\label{tubefig}
\end{figure}

Although we have discussed theoretical aspects of the $S=1/2$ isosceles quantum spin tube, 
a compound corresponding to our model has not been found yet. 
In this section, we review the twisted spin tube, which is of particular importance for actual experiments;
Recently, the interesting compound [(CuCl$_2$tachH)$_3$Cl]Cl$_2$ was actually synthesized as an assembly of the triangular cluster conformed by Cu$^{++}$ ions  and  magnetization measurements were performed \cite{nojiri}.
As is in figure \ref{tubefig},  the tube structure of this compound is based on the alternatingly-aligned triangles, where the exchange coupling in the unit triangle is denoted as $J_{\rm r}$ and the intra-triangle coupling is written as $J_1$.
The low-energy properties of the quantum spin tube were theoretically studied by DMRG \cite{okunishi,fouet}.
Also,  the quantum spin tube with the easy-plane anisotropy was investigated in the context of the triangular lattice spin system  and then interesting field induced phase transitions were observed in the magnetization curve \cite{yoshi}.

At the early stage of study, [(CuCl$_2$tachH)$_3$Cl]Cl$_2$ might have a small spin gap, based on the susceptibility measurement above 2K. 
However, the succeeding studies have  clarified that the observed spin-gap-like behavior around $2$K crossovers to the gapless behavior at very low temperature region ($T<0.5$K) \cite{nojiriarxive}, which is consistent with the theoretical results.
This suggests that a kind of dual structure associated with the quantum phase transition of the first order may be captured by the experimental measurements.
Below, we will review the quantum phase transition peculiar to the twisted quantum spin tube.

The model Hamiltonian of the twisted spin tube is written as,
\begin{equation}
\label{twistedH}
\cH= J_1 \sum _{i=1}^3 \sum_{j=1}^L \left[ \vS_{i,j}\cdot \vS_{i,j+1}+\vS_{i,j}\cdot \vS_{i+1,j+1}\right]
   +J_{\rm r} \sum _{i=1}^2 \sum_{j=1}^L \vS_{i,j}\cdot \vS_{i+1,j} 
\end{equation}
with the periodic boundary condition in the $i$-direction. 
A naive consideration about the lattice structure of this twisted tube suggests that a quantum phase transition between the following two phases may occur; 
For $J_{\rm r}\gg J_1$, the system is described by the almost decoupled triangles, where the chirality degree of freedom defined on the unit triangle plays crucial role,  while for $J_{\rm r}\ll J_1$, the lattice structure is basically described as a rhombus lattice. 
Note that, if $J_{\rm }=0$, the rhombus lattice has no frustration.
Thus we can expect a quantum phase transition between  decoupled-triangles and rhombus lattice phases.

In the decoupled triangle limit($J_{\rm r}\gg J_1$), the degenerate perturbation with respect to $J_1$ leads us to the effective spin-chirality model,
\begin{equation}
\fl
   {\cal H}_{\rm eff}
   = \frac{2J_1}{3}\sum_j \vT_j\cdot\vT_{j+1}
     [ 1 + 2 (\exp(\rmi \pi/3) \tau^+_j\tau^-_{j+1} +(\exp(-\rmi \pi/3)\tau^-_j\tau^+_{j+1})]
     \label{twistedspinchiral}
\end{equation}
where $\vT$ represents the $S=1/2$ spin operator and $\tau$ indicates an effective spin matrix representing the chirality degrees of freedom as in (\ref{eq:Heff-for-regular}).
This effective Hamiltonian is very similar to (\ref{eq:Heff-for-regular});
The phase factor $e^{i\pi/3}$ originates from the $\pi/3$ rotation of the lattice along the tube direction and it can be removed by the gauge transformation. 
On the same line of the argument as (\ref{eq:Heff-for-regular}), the effective Hamiltonian (\ref{twistedspinchiral}) has a spin gap\cite{schulz,kawano,mila}.
Indeed,  a detailed DMRG computation of the full Hamiltonian (\ref{twistedH}) actually confirms that the spin gap exists for $J_1/J_{\rm r} < 1.21$ \cite{okunishi,fouet}.
For $J_1/J_{\rm r}>1.21$, on the other hand, the finite size scaling analysis of the spin gap basically indicates the gapless ground state \cite{okunishi}.
Although the usual singlet-triplet spin gap for the open boundary condition captures a boundary excitation in $1.21<J_1/J_{\rm r}<1.5$,  such an anomalous behavior due to the boundary can be settled down by using the single-spin termination of the tube \cite{okunishi2}.

In order to resolve nature of the quantum phase transition of the twisted tube, an important quantity is the total-$S$ on the unit triangle \cite{fouet}. 
For decoupled triangle, the ground state is in doubly degenerating doublets of the unit triangle, while for $J_{\rm r}=0$, the ground state basically belongs to $S=3/2$ sector.  
Here, we define the projection operator of the total-$S$ of the unit triangle to the doublet sector  as $ P_{1/2}( \equiv \frac{1}{2} -\frac{2}{3}(\vS_1\cdot \vS_2 +\vS_2\cdot \vS_3+  \vS_3\cdot \vS_1 ))$ for each unit triangle.
As discussed in \cite{fouet}, then, this projection operator $P_{1/2}$ is a good order parameter of the twisted tube; The detailed computation of the expectation value $\langle P_{1/2} \rangle$ shows discontinuity
at the critical value $(J_1/J_{\rm r})_c \simeq1.22$ not at 
$J_1/J_{\rm r} \simeq1.47$,
which is consistent with the spin gap result.
In addition to this, we should remark that an extended spin tube with the diagonal interaction shows
 the exact first order phase transition, for which total-$S$ on the unit triangles independently conserves \cite{honecker}.
We have thus investigated how to connect the twisted tube to the diagonal interaction model. 
We add
\begin{equation}
{\cal H}_{\gamma J_1}= \gamma J_1\sum _{i=1}^3 \sum_{j=1}^L  \vS_{i,j}\cdot \vS_{i-1,j+1}
\end{equation}
to the Hamiltonian (\ref{twistedH}), and then varies $ -1 \le \gamma \le 1$.
Note that $\gamma=0$ corresponds to the original twisted tube and $\gamma=1$ does to the diagonal spin tube, which shows the first order transition at $J_1/J_{\rm r} \simeq 0.63$\cite{honecker}.

\begin{figure}
\begin{center}
\rotatebox{-90}{\scalebox{0.4}{\includegraphics{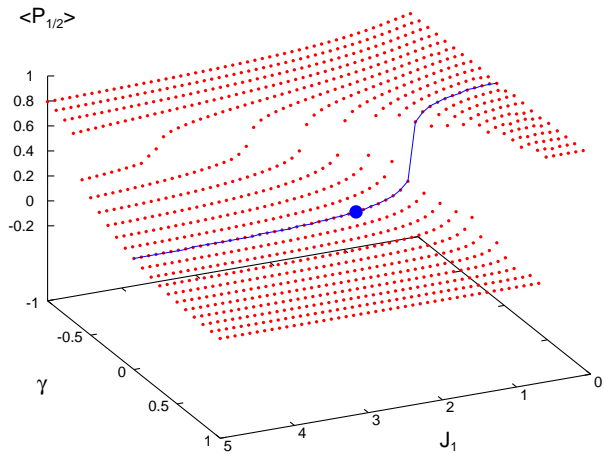}}}
\end{center}
\caption{(color online)
 The expectation value of the projection operator $\langle P_{1/2} \rangle $ for the extended quantum spin tube.
The solid line represents  $\gamma=0$ and the solid circle indicates the parameters corresponding to [(CuCl$_2$tachH)$_3$Cl]Cl$_2$. }
\label{twistPD}
\end{figure}

In figure \ref{twistPD}, we show $\langle P_{1/2} \rangle $ at the center of the extended tube calculated by DMRG for $L=36$ with $J_{\rm r}=1$.
In the figure, $\langle P_{1/2}\rangle $  jumps exactly from  zero to unity at $J_1/J_{\rm r}\simeq 0.63 $ along the line of $\gamma=1$, where the total-$S$ of the unit triangle is the exact symmetry. 
Then, we can see that the ``gap'' of  the $\langle P_{1/2} \rangle $ is adiabatically continued to $\gamma=0$.
This implies that the transition at $(J_1/J_{\rm r})_c\simeq 1.21$ on the line of $\gamma=0$ corresponding the twisted tube (\ref{twistedH}) is of first order. 
As $J_1$ increases in  $\gamma <0$, this gap reduces, but we can not confirm whether the end point of the first order transition exist or not within the present calculations for $L=36$.
However, we think that the overview of the first order quantum phase transition is clarified.

In connection with the experiments,  an interesting point is that the parameter $J_1/J_{\rm r} \simeq 2.16$ for  [(CuCl$_2$tachH)$_3$Cl]Cl$_2$ is located in the gapless  $S=3/2$ sector, but it is surrounded by the ``gap'' of the first-order transition, as in figure \ref{twistPD}.
This implies that, although the ground state itself is gapless,  a certain density of states originating from the $S=1/2$ sector can be expected above the ground state.
Thus we can expect that such dual structure of $S=1/2$ and 3/2 sectors affects the experimentally observed quantities of the twisted tube at a finite temperature.
Indeed, a recent specific heat result illustrates that the linear temperature dependence at very low temperature($T<0.5$K), which can be described as an effective $S=3/2$ chain,  crossovers to spin-gap-like increasing around $T\sim 2$K \cite{nojiriarxive}.
Also for the low-field magnetization curve of $J_1/J_{\rm r}>1.21$, 
we can see that, after linear increase of the magnetization at very low field region, the slope of the magnetization curve rapidly increases, as if it had a spin gap \cite{okunishi}.

\section{Future prospects}

As a future prospect, it would be interesting to consider the carrier-doped 
spin nanotube like the high-Tc cuprates, where the system is effectively 
described by the Hubbard model near the half-filling case. 
A mechanism of the superconductivity based on the spin gap had been proposed 
for the carrier-doped two-leg spin ladder \cite{rice} 
and actually a pressure-induced 
superconductivity was observed on the spin ladder cuprate \cite{uehara}. 
Motivated by the discovery, the spin gap mediated superconductivity was 
theoretically proposed even for the three-leg spin ladder, using the 
quantum Monte Carlo simulation \cite{kimura1,kimura2}. 
If the carrier-doped three-leg spin tube is realized, it would be 
a better candidate of the superconductor, rather than the three-leg ladder, 
because it has a spin gap. 
In addition the three-leg spin tube was also revealed to have an energy gap 
in the chirality degrees of freedom \cite{mila}. 
It suggests that a chirality induced superconductivity would be possibly 
realized as a new mechanism of superconductivity in the near future.

\ack

We thank Dr. Yuichi Ohtsuka for co-working
in the initial stage of this work.
We also thank Profs. I Affleck, A. L\"auchli, C. Lhuillier, 
H. Manaka, F. Mila, H. Nojiri, D. Poilblanc, P. Pujol, 
J. Schnack, P. Sindzingre, and Drs. D. Charrier, 
G. N\'enert for fruitful discussions. 
This work has been
partly supported by Grants-in-Aid for Scientific Research (B) 
(No.17340100, No.20340096), Scientific Research (C) 
(No.18540340),
for Young Scientists (B) (No.21740295),
and 
on Priority Areas ``Invention of Anomalous Quantum Materials --- New 
Physics through Innovation Materials ---'' (No.19014019), 
``Physics of New Quantum Phases in Superclean Materials'' 
(No.17071011, No.18043023, No.20029020),
 ``High Field Spin Science in 100T'' (No.20030008, No.2003003)
and
``Novel States of Matter Induced by Frustration'' (No.22014016, No.22014012)
from the Ministry
of Education, Culture, Sports, Science and Technology of Japan.
We further thank the Supercomputer Center, Institute
for Solid State Physics, University of Tokyo, the Cyberscience Center, Tohoku
University, and the Computer Room, Yukawa Institute for Theoretical Physics,
Kyoto University for computational facilities.


\section*{References}

\begin{thebibliography}{9}

\bibitem{seeber}
Seeber G, K\"ogerler P, Kariuki B M and Cronin L, Chem. Commun. (2004) 1580

\bibitem{nojiri}
Schnack J, Nojiri H, K\"ogerler P, Cooper G J T and Cronin L 2004 
\PR B {\bf 70} 174420 

\bibitem{manaka}
Manaka H, Hirai Y,  Hachigo Y, Mitsunaga M, Ito M and
Terada N 2009  
{\it J. Phys. Soc. Jpn.} {\bf 78} 093701 

\bibitem{millet}
Millet P, Henry J Y, Mila F and  Galy J 1999 
{\it J. Solid State Chem.} {\bf 147} 676

\bibitem{garlea}
Garlea V O, Zheludev A, Regnault L -P, Chung J -H, Qiu Y,
Boehm M, Habicht K and Meissner M 2008
{\it Phys. Rev. Lett.} {\bf 100} 037206

\bibitem{zheludev}
Garlea V O, Zheludev A, Habicht K, Meissner M, Grenier B,
Regnault L -P and Ressouche E 2008 arXiv:0807.1571. 

\bibitem{sakai2}
Sakai T, Sato M, Okunishi K, Otsuka Y, Okamoto K and Itoi C 2008  
\PR B {\bf 78} 184415

\bibitem{okunishi2}
Okunishi K, Yoshikawa S, Sakai T and Miyashita S 2009 
{\it Int. J. Mod. Phys.} C {\bf 20} 1423 

\bibitem{sakai4}
Sakai T, Okunishi K, Okamoto K, Sato M, Matsumoto M and Otsuka Y 2007 
{\it J. Magn. Magn. Mater.} {\bf 310} e423 

\bibitem{schulz}
Schulz H J 1996 in {\it Correlated Fermions and Transport in Mesoscopic Systems}, 
ed. Martin T, Montambaux G, Tran Than Van J (cond-mat/9605075) 

\bibitem{kawano}
Kawano K and Takahashi M 1997 
{\it J. Phys. Soc. Jpn.} {\bf 66} 4001
 
\bibitem{cabra}
Cabra D C, Honecker A and Pujol P 1998 
\PR B {\bf 58} 6241

\bibitem{mila}
Luscher A, Noack R M, Misguich G, Kotov V N and Mila F 2004 
\PR B {\bf 70}  060405(R) 

\bibitem{fouet}
Fouet J -B, L\"auchli A, Pilgram A, Noak R M and Mila F 2006, 
\PR B {\bf 73} 014409

\bibitem{okunishi}
Okunishi K, Yoshikawa S, Sakai T and Miyashita S 2005 
{\it Prog. Theor. Phys.} Suppl. {\bf 159} 297

\bibitem{sakai}
Sakai T, Matsumoto M, Okunishi K, Okamoto K and Sato M 2005 
{\it Physica} E {\bf 29} 633 

\bibitem{arikawa}
Nishimoto S and Arikawa M 2008 \PR B {\bf 78} 054421  

\bibitem{sato07}
Sato M 2007 \PR B {\bf 75} 174407

\bibitem{sato-sakai}
Sato M and Sakai T 2007 \PR B {\bf 75} 014411

\bibitem{sato05}
Sato M 2005 \PR B {\bf 72} 104438

\bibitem{sato-oshikawa}
Sato M and Oshikawa M 2007 \PR B {\bf 75} 014404

\bibitem{lieb}
Lieb E, Schultz T and Mattis D 1961 {\it Ann. Phys.} {\bf 16} 407 


\bibitem{sakai3}
Sakai T, Okunishi K, Okamoto K, Itoi C and Sato M 2010 
{\it J. Low Temp. Phys.} {\bf 159}  55 

\bibitem{matsumoto}
Matsumoto M, Sakai T, Sato M, Takayama H and Todo S 2005 
{\it Physica } E {\bf 29} 660 

\bibitem{charrier}
Charrier D, Caopponi S, Oshikawa M and Pujol P 2010 
cond-mat/1005.0711. 

\bibitem{affleck2}
Affleck I, Nucl. Phys. B {\bf 265} (1986) 409

\bibitem{ledermann}
Ledermann U, Le Hur K and Rice T M, Phys. Rev. B {\bf 62} (2000) 16383

\bibitem{Aff-Hal}
Affleck I and Haldane F D 1987 \PR B {\bf 36} 5291

\bibitem{balents}
Balents L and Fisher M P 1996 \PR B {\bf 53} 12133

\bibitem{Arrigoni}
Arrigoni E 1996 {\it Phys. Lett.} A {\bf 215} 91

\bibitem{lin}
Lin H, Balents L and Fisher M P 1997 
\PR B {\bf 56} 6569

\bibitem{affleck}
Affleck I 1989 in {\it Fields, Strings and Critical Phenomena, 
	1988 Les Houches Lecture Notes}, edited by E. Br\'ezin and
	J. Zinn-Justin (Elsevier, Amsterdam) p~564.

\bibitem{gogolin}
Gogolin A O, Nersesyan A A and Tsvelik A M 1998
{\it Bosonization and Strongly Correlated Systems}
(Cambridge University Press, UK)

\bibitem{tsvelik}
Tsvelik A M 2003 {\it Quantum Field Theory in Condensed Matter
	Physics, 2nd edition} (Cambridge University Press, UK)

\bibitem{totsuka}
Totsuka K and Suzuki M 1996 \JPA {\bf 29} 3559

\bibitem{bere}
Berezinskii Z L 1970 {\it Zh. Eksp. Teor. Fiz.} {\bf 59} 907 
({\it Sov. Phys.}-JETP 1971 {\bf 32} 493)
 
\bibitem{kt}
Kosterlitz J M and Thouless D J 1973 \JPC {\bf 6} 1181 

\bibitem{Note1} 
We have neglected the difference between the Fermi velocities of the
	first and second bands. 

\bibitem{shelton}
Shelton D G, Nersesyan A A and Tsvelik A M 1996 \PR B 
{\bf 53} 8521


\bibitem{on}
Okamoto K and Nomura K 1992 {\it Phys. Lett.} A {\bf 169} 433 

\bibitem{cardy2}
Cardy J L 1986 \JPA {\bf 19} L1093

\bibitem{cardy3}
Cardy J L 1987 \JPA {\bf 20} 5039

\bibitem{no}
Nomura K and K. Okamoto K 1994 \JPA {\bf 27} 5773

\bibitem{LS2} 
Okamoto K 2002 {\it Prog. Theor. Phys,} Suppl. No. {\bf 145} 113

\bibitem{LS3}
Nomura K and Kitazawa A 2002 {\it Proc. French-Japanese
 Symp. on Quantum Properties of Low-Dimensional
 Antiferromagnets} ed Y. Ajiro and J-P. Bouhcer (Kyushuu
 University Press, Fukuoka, Japan) (cond-mat/020172)

\bibitem{cft1}
Cardy J L 1984 \JPA {\bf 17} L385 

\bibitem{cft2}
Bl\"ote H W, Cardy J L and Nightingale M P 1986
{\it Phys. Rev. Lett.} {\bf 56}  742 

\bibitem{cft3}
Affleck I 1986 {\it Phys. Rev. Lett.} {\bf 56} 746 

\bibitem{IK} 
Itoi C and  Kato M -H 1997 \PR B {\bf 55} 8295

\bibitem{GS} 
Giamarchi T and Schulz H J 1989 \PR B {\bf 39} 4620

\bibitem{ZS} 
Ziman T and Schulz H J 1987 {\it Phys. Rev. Lett.} {\bf 59} 140 

\bibitem{giamarchi}
Giamarchi T 2004 {\it Quantum Physics in One Dimension} 
(Oxford University Press)

\bibitem{fabrizio}
Fabrizio M, Gogolin A O and Nersesyan A A 2000
\NP B{\bf 580} 647

\bibitem{bajnok}
Bajnok Z, Palla L , Tak\'acs G, and  W\'agner F 2001 
\NP B{\bf 601} 503

\bibitem{fath}
F\'ath G 2003 \PR B {\bf 68} 134445

\bibitem{essler04}
Essler  F H L and Affleck I 2004 
{\it J. Stat. Mech.: Theor. Exp.} P12006


\bibitem{sato06}
Sato M 2006 {\it J. Stat. Mech.: Theor. Exp.} P09001

\bibitem{okamotoetal} Okamoto K,
Sakai T, Sato M, Okunishi K and Itoi C, in preparation

\bibitem{dmitriev1}
Dmitriev D V, Krivnov V Ya and Ovchinnikov 2002
\PR B {\bf 65}, 172409

\bibitem{dmitriev2}
Dmitriev D V, Krivnov V Ya, Ovchinnikov and Langari A 2002
{\it J. Exp. Theor. Phys.} {\bf 95} 538
({\it Zh. Eksp. Teor. Fiz.} {\bf 122} 624)

\bibitem{dmitriev3}
Dmitriev D V and Krivnov 2004
\PR B {\bf 70} 144414

\bibitem{capraro}
Capraro F and Gros C 2002
{\it Eur. Phys. J.} B {\bf 29} 35

\bibitem{tit-gcoe}
Okamoto K, Sato M, Okunishi K, Sakai T and Itoi C 2010
{\it Physica E} to appear

\bibitem{cross-fisher}
Cross M and Fisher D S 1979 \PR B {\bf 19} 402

\bibitem{nakano-fukuyama}
Nakano T and Fukuyama H 1980 {\it J. Phys. Soc. Jpn.} {\bf 49} 1679

\bibitem{ont}
Okamoto K, Nishimori H and Taguchi Y 1986
{\it J. Phys. Soc. Jpn.} {\bf 55} 1458


\bibitem{black-emery}
Black J L and Emery V J 1981 \PR B {\bf 23} 429

\bibitem{yoshi}
Yoshikawa S, Okunishi K, Senda M and Miyashita S 2004
{\it J. Phys. Soc. Jpn.} {\bf 73} 1798


\bibitem{nojiriarxive}
Ivanov N B, Schnack J, Schnalle R, Richter J, K\"ogerler P, Newton G N, 
Cronin G, Oshima Y and Nojiri H, Phys. Rev. Lett. 105 (2010) 037206

\bibitem{honecker}
Honecker A, Mila F and Troyer M 2000, {\it Eur. Phys. J.} B {\bf 15} 227

\bibitem{rice}
Rice M, Gopalan S and Sigrist M 1993
{\it Europhys. Lett.} {\bf 23} 445 

\bibitem{uehara}
Uehara M, Nagata T, Akimitsu J, Takahashi H, M\^ori N
and Kinoshita K 1996 
{\it J. Phys. Soc. Jpn.} {\bf 65} 2764 

\bibitem{kimura1}
Kimura T, Kuroki K and Aoki H 1996 \PR B {\bf 54} R9608 

\bibitem{kimura2}
Kimura T, Kuroki K and Aoki H 1998 {\it J. Phys. Soc. Jpn.} {\bf 67} 
1377 


\end{thebibliography}
\end{document}